\documentclass[lettersize,journal]{IEEEtran}
\usepackage{amsmath,amsfonts, amssymb}
\usepackage{bbm}
\usepackage{algorithmic}
\usepackage{algorithm}
\usepackage{array}
\usepackage{subcaption}
\usepackage{textcomp}
\usepackage{stfloats}
\usepackage{url}
\usepackage{xcolor}
\usepackage{physics}
\usepackage{verbatim}
\usepackage{graphicx}
\usepackage{cite}

\newcommand{\rev}{\color{black}}

\newtheorem{theorem}{Theorem}

\newtheorem{lemma}{Lemma}
\newtheorem{proposition}{Proposition}
\newtheorem{remark}{Remark}
\newtheorem{example}{Example}
\newtheorem{definition}{Definition}

\begin{document}

\title{Enhancing Sum Capacity via Quantum and \\ No-Signaling Cooperation Between Transmitters}

\author{Seung-Hyun Nam,~\IEEEmembership{Member,~IEEE,} Hyun-Young Park,~\IEEEmembership{Member,~IEEE,} Jiyoung Yun, \\ Ashutosh Rai, Si-Hyeon Lee,~\IEEEmembership{Senior Member,~IEEE,} and Joonwoo Bae,~\IEEEmembership{Member,~IEEE}

\thanks{© 2026 IEEE. Personal use of this material is permitted. Permission from IEEE must be obtained for all other uses, in any current or future media, including reprinting/republishing this material for advertising or promotional purposes, creating new collective works, for resale or redistribution to servers or lists, or reuse of any copyrighted component of this work in other works.}
\thanks{S.-H. Nam and J. Yun are with the Information \& Electronics Research Institute, Korea Advanced Institute of Science and Technology (KAIST), Daejeon 34141, South Korea (e-mails: shnam@kaist.ac.kr; jiyoungyun@kaist.ac.kr).}
\thanks{H.-Y. Park, A. Rai, S.-H. Lee, and J. Bae are with the School of Electrical Engineering, KAIST, Daejeon 34141, South Korea  (e-mails: phy811@kaist.ac.kr;  ashutosh.rai@kaist.ac.kr; sihyeon@kaist.ac.kr; joonwoo.bae@kaist.ac.kr). (Corresponding Author: Si-Hyeon Lee and Joonwoo Bae).
}
}

\maketitle

\begin{abstract}
We consider communication over discrete memoryless interference channels or multiple access channels without feedback, where transmitters exploit classical, quantum, or no-signaling cooperation.
{\rev Previous works have shown that, for channels associated with pseudo-telepathy games, quantum or no-signaling cooperation can increase the sum capacity.
However, a full characterization of channels admitting such an improvement remains open.
Motivated by common features of previously studied examples, we propose a broader class of game-induced channels.
In these channels, each input is a question–answer pair from a nonlocal game.
When the inputs satisfy the game’s winning condition, the channel has lower conditional output uncertainty, and the channel decomposes into parallel weakly symmetric subchannels with a unique capacity-achieving input distribution.
We show that, for this class, quantum or no-signaling cooperation strictly increases the sum capacity whenever the associated game is a quantum or no-signaling pseudo-telepathy game, respectively.
The proposed class recovers several previously studied channels and includes new examples.
}
\end{abstract}


\section{Introduction}
\IEEEPARstart{Q}{uantum} information processing has attracted significant attention due to its potential to outperform classical information processing in certain tasks, thus exhibiting quantum advantage.
Quantum advantage is not only of theoretical interest but also provides justification for the practical implementation of quantum information processing.
Examples of quantum advantage arise in the contexts of algorithmic complexity \cite{shorPolynomialTimeAlgorithmsPrime1997, groverFastQuantumMechanical1996, deutsch_rapid_1992}, security \cite{bennettQuantumCryptographyPublic2014, ekertQuantumCrypt1991,bennett_quantum_1992}, privacy \cite{namQuantumAdvantagePrivate2025}, nonlocal games \cite{brassardQuantumPseudoTelepathy2005a,cleveConsequencesLimitsNonlocal2004, brunnerBellNonlocality2014}, distributed computing \cite{cleve_substituting_1997, bar-yossef_exponential_2004, buhrmanNonlocalityCommunicationComplexity2010a}, and communication over networks \cite{leditzkyPlayingGamesMultiple2020, hawellekInterferenceChannelEntangled2025, peregMultipleAccessChannelEntangled2025, seshadriSeparationCorrelationAssistedSum2023, YunRaiBae2020, 
yunNonLocalQuantumAdvantages2023}.

{\rev In communication over networks, previous works \cite{leditzkyPlayingGamesMultiple2020, hawellekInterferenceChannelEntangled2025, peregMultipleAccessChannelEntangled2025, YunRaiBae2020, seshadriSeparationCorrelationAssistedSum2023, 
yunNonLocalQuantumAdvantages2023} considered scenarios in which multiple transmitters use pre-shared resources to cooperate while sending classical messages over classical interference channels (ICs) or multiple access channels (MACs).
For a given IC or MAC, the fundamental limit of communication is characterized by the capacity region, i.e., the set of all achievable rate tuples. In this setting, a quantum advantage is established when quantum cooperation enlarges the capacity region compared to classical cooperation. Such an advantage can also be demonstrated through the sum capacity, defined as the maximum achievable sum rate, if quantum cooperation leads to a larger sum capacity. 
}

\begin{figure}
    \centering
    \includegraphics[width=0.4\linewidth]{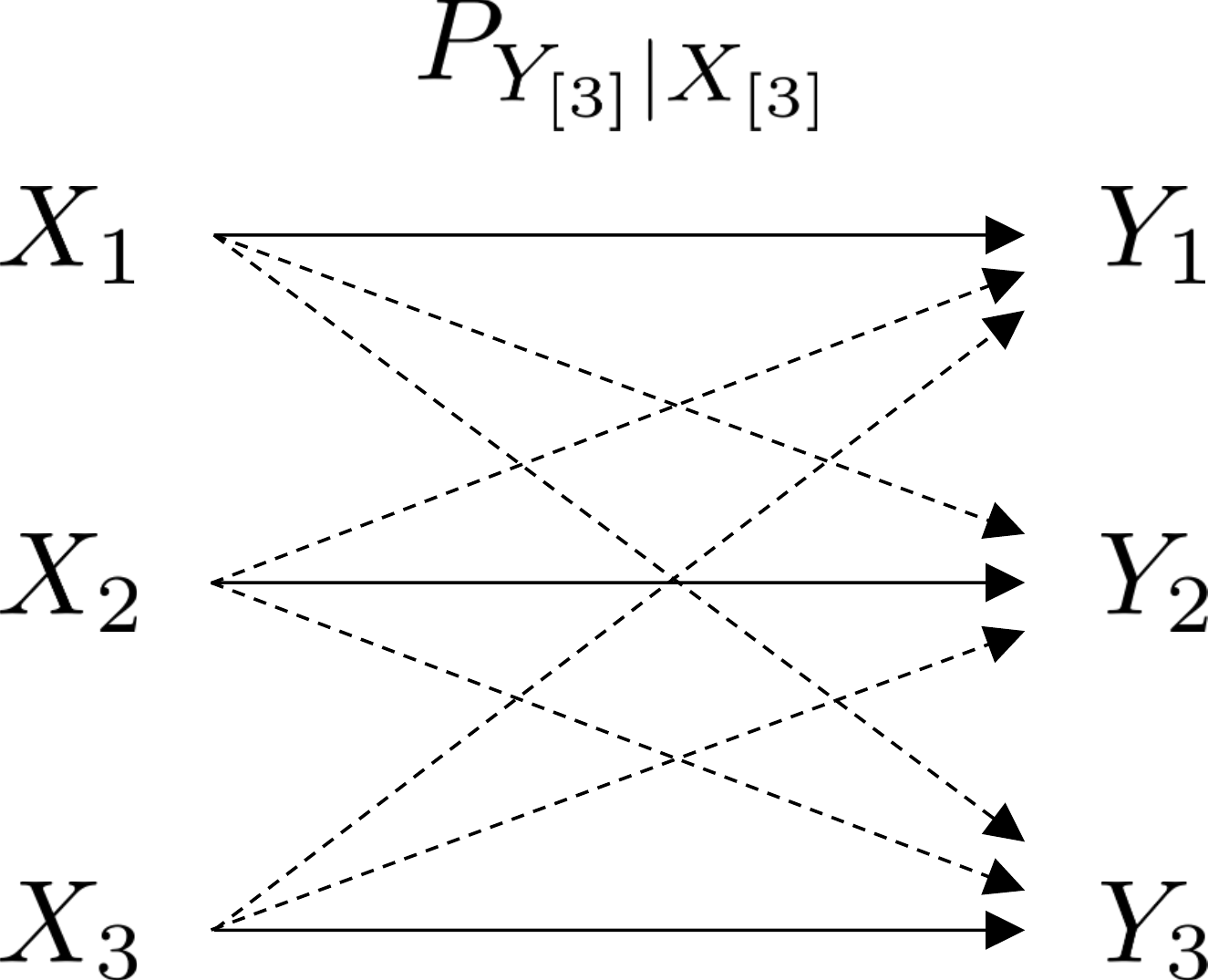}
    \caption{{\rev A conceptual diagram for an interference channel $P_{Y_{[K]}|X_{[K]}}$ between $K=3$ transmitter-receiver pairs. Channel output $Y_i$ at receiver $i$ can be affected by unintended channel input $X_{i'}$ from transmitter $i'$, $i' \neq i$ (dashed lines).
    Correlations between channel inputs can mitigate cross-link dependencies.
    }}
    \label{fig:IC}
\end{figure}

{\rev Intuitively, quantum advantage arises from non-classical correlations induced by shared quantum resources.
In both ICs and MACs, transmitted signals from multiple transmitters interfere with each other, thereby making it more difficult for a receiver to decode the desired message encoded in the transmitted signal, as depicted in Figure~\ref{fig:IC} for an IC.
Within this scenario, transmitters may cooperate with each other to generate correlated signals that would mitigate interference and increase the sum capacity.
At this point, transmitters sharing a quantum resource can generate more general correlations than sharing a classical resource, and can thereby mitigate more general types of interference.
Note that certain quantum correlations cannot be reproduced by local classical systems under spacelike separation, yet they remain constrained by physical principles such as relativistic causality \cite{popescuQuantumNonlocalityAxiom1994}.
This constraint is captured by no-signaling correlations, which require that the marginal statistics at each local site be independent of the local operations performed at the other sites.
No-signaling cooperation has also been considered as a means of increasing the sum capacity \cite{quekQuantumSuperquantumEnhancements2017, YunRaiBae2020, yunNonLocalQuantumAdvantages2023, seshadriSeparationCorrelationAssistedSum2023}.
Thus, it is natural to ask \emph{for which class of channels the sum capacity can be strictly enlarged by quantum or no-signaling cooperation, compared with classical correlation}.
}

{\rev Regarding this question, previous works \cite{YunRaiBae2020, leditzkyPlayingGamesMultiple2020, hawellekInterferenceChannelEntangled2025, peregMultipleAccessChannelEntangled2025, seshadriSeparationCorrelationAssistedSum2023, 
quekQuantumSuperquantumEnhancements2017, yunNonLocalQuantumAdvantages2023} demonstrated sum rate enhancement by quantum or no-signaling cooperation for some specific channels involving pseudo-telepathy games.} 
A pseudo-telepathy game is a nonlocal game whose winning probability is strictly less than one with classical cooperation but equal to one with quantum or no-signaling cooperation.
Examples of pseudo-telepathy games include the Clauser-Horne-Shimony-Holt (CHSH) game \cite{clauserProposedExperimentTest1969}, the magic square game \cite{aravindBellsTheoremInequalities2002}, and the multiparty parity game \cite{brassardQuantumPseudoTelepathy2005a}.
These games exhibit a hierarchy among three types of cooperation: quantum cooperation produces correlations unattainable by classical cooperation, and no-signaling cooperation generates correlations unattainable by quantum cooperation.
Interpreted in the context of communication, quantum or no-signaling cooperation between transmitters can generate correlated channel inputs that are not possible through classical cooperation.
Consequently, pseudo-telepathy games give rise to network channels in which interference between channel inputs can be mitigated by quantum or no-signaling cooperation, but not by classical cooperation.

Specifically, the works \cite{YunRaiBae2020, leditzkyPlayingGamesMultiple2020, hawellekInterferenceChannelEntangled2025, peregMultipleAccessChannelEntangled2025, seshadriSeparationCorrelationAssistedSum2023, 
quekQuantumSuperquantumEnhancements2017, yunNonLocalQuantumAdvantages2023} defined discrete memoryless MACs and ICs where channel inputs are specified as tuples of questions and answers from a pseudo-telepathy game.
{\rev Here, a question can be interpreted as the message-bearing component of the transmitted signal, while an answer can be viewed as a finite-alphabet coordination variable, analogous to a precoding or beamforming choice.}
Additionally, the channels studied in \cite{leditzkyPlayingGamesMultiple2020, hawellekInterferenceChannelEntangled2025, peregMultipleAccessChannelEntangled2025, seshadriSeparationCorrelationAssistedSum2023,quekQuantumSuperquantumEnhancements2017} relay the questions from transmitters to one or more receivers noiselessly when the channel inputs satisfy the winning condition of the game.
More general channels have also been considered, where the transmitted questions are subject to random erasure \cite{quekQuantumSuperquantumEnhancements2017, YunRaiBae2020}, or to uniform random noise \cite{yunNonLocalQuantumAdvantages2023}, even when the channel inputs satisfy the winning condition.
However, it has not yet been fully determined for which channels the sum capacity can be increased by quantum or no-signaling cooperation.
{\rev For example, we may consider more general types of noise other than erasure or uniform random noise.}

In this work, we propose a broader class of channels by focusing on the common characteristics shared by the previously studied channels: \emph{the interference is weaker when the channel inputs satisfy the winning condition of a pseudo-telepathy game than when they do not.}
We then show the enhancement of the sum capacity through quantum or no-signaling cooperation for this class of channels.
Specifically, we propose a class of channels involving pseudo-telepathy games {\rev by interpreting \emph{weaker interference} in the following manner: When the channel inputs satisfy the winning condition of the game, a channel has lower conditional output uncertainty in terms of the conditional entropy of the channel output given channel inputs, and decomposes into parallel weakly symmetric subchannels with unique capacity-achieving input distribution.
By establishing these sufficient conditions for the enhancement of the sum capacity derived from structural properties, rather than investigating specific channels case-by-case, our framework can be generalized to a broad class of interference channels.

Decomposition into parallel subchannels means that, in the winning state, each receiver observes a signal that depends only on its corresponding transmitter's question and not on the questions of the other transmitters. Thus, the winning condition acts as an interference-alignment condition: successful coordination removes cross-link interference, leaving only local point-to-point noise.
The weak-symmetry condition yields a closed-form capacity expression for each winning-state subchannel, while uniqueness of the capacity-achieving input distribution is used in the converse proof to show that the sum capacity upper bound cannot be attained under classical cooperation.
Moreover, this assumption holds for simple and physically motivated examples, including uniform random noise channels and cyclic nearest-neighbor confusion channels (Examples~\ref{ex:unif_ch} and~\ref{ex:nearest_neighbor}).
} 

The proposed class of channels includes the channels studied in prior works \cite{leditzkyPlayingGamesMultiple2020, seshadriSeparationCorrelationAssistedSum2023, hawellekInterferenceChannelEntangled2025, peregMultipleAccessChannelEntangled2025} as well as new ones not considered in \cite{YunRaiBae2020, leditzkyPlayingGamesMultiple2020, hawellekInterferenceChannelEntangled2025, peregMultipleAccessChannelEntangled2025, seshadriSeparationCorrelationAssistedSum2023, 
quekQuantumSuperquantumEnhancements2017, yunNonLocalQuantumAdvantages2023}.
To show the enhancement, we first derive an upper bound on the sum capacity of the proposed channels, which cannot be achieved by classical cooperation.
Next, we show that the sum rate equal to this upper bound can be achieved when the transmitters exploit cooperation that wins the associated pseudo-telepathy game with probability one.

{\rev
We remark that while there have also been studies exploring more general scenarios involving quantum or no-signaling cooperation among both transmitters and receivers, as well as network channel models beyond 
MACs and ICs \cite{fawziBroadcastChannelCoding2024, fawziMultipleAccessChannelCoding2024, notzelEntanglementEnabledCommunication2020}, we focus on MACs and ICs with quantum 
and no-signaling cooperation between transmitters, as considered in \cite{YunRaiBae2020, leditzkyPlayingGamesMultiple2020, hawellekInterferenceChannelEntangled2025, peregMultipleAccessChannelEntangled2025, seshadriSeparationCorrelationAssistedSum2023, 
quekQuantumSuperquantumEnhancements2017, yunNonLocalQuantumAdvantages2023}.
Even in these simpler settings, it remains an open question under which classes of channels such an improvement can be achieved.
}

The rest of this paper is organized as follows. 
In Section II, we introduce the communication scenario with and without cooperation between transmitters, as well as preliminaries on nonlocal games and the hierarchy of cooperation.
In Section III, we propose a class of channels involving nonlocal games and prove the enhancement of sum capacity via quantum or no-signaling cooperation.
Section IV concludes the paper with a discussion of future work.

\section{System model and Preliminaries}
In this work, we propose a class of ICs and MACs whose sum capacities can be enhanced through quantum or no-signaling cooperation between transmitters.
Since all results for ICs can be directly adapted to those for MACs by treating multiple receivers as a single receiver (see Remark~\ref{rmk:IC_to_MAC} for details), we primarily focus on ICs throughout this paper.
To illustrate our results, we begin with a review of the conventional communication scenario over an IC, in which transmitters and receivers do not cooperate.
Then, we describe an extended scenario that allows cooperation between transmitters, which is the focus of this work.
Finally, we introduce nonlocal games, which serve as the key component underlying the proposed class of channels.

\vspace{10pt}

\emph{Notation}:
For $K \in \mathbb{N}$, let $[K] := \{1,\ldots,K\}$.
For finite sets $\mathcal{X}_1,\mathcal{X}_2,\ldots,\mathcal{X}_K$, we denote $\mathcal{X}_{[K]}:= \prod_{i\in[K]} \mathcal{X}_i$, $\mathcal{X}_i^n:= \prod_{j \in [n]}\mathcal{X}_{i,j}$ where $\mathcal{X}_{i,j} = \mathcal{X}_i$ for all $j \in [n]$, and  $\mathcal{X}_{[K]}^n:= \prod_{j\in [n]} \mathcal{X}_{[K],j}$ where $\mathcal{X}_{[K],j}:= \mathcal{X}_{[K]}$.
A constant (e.g. $x_{[K]}^n$) or a random variable (e.g. $X_{[K]}^n$) with a given index is regarded as belonging to the corresponding set (e.g. $\mathcal{X}_{[K]}^n$).
{\rev Throughout the paper, for blocklength-$n$ channel input and output variables, the first subscript denotes the transmitter–receiver pair, while the second subscript denotes the channel-use index.
For example, $X_{i,j}$ denotes the channel input of transmitter $i$ at the $j$-th channel use.
Also, $X_{i}^n$ denotes the channel inputs of transmitter $i$ over $n$ channel uses, and $X_{[K]}^n$ denotes the channel inputs of all transmitters $1$ to $K$ over $n$ channel uses.
}

\subsection{Communication over IC without transmitter cooperation}\label{sec:DMIC}
The conventional communication scenario between $K$ transmitter-receiver pairs, $K \geq 2$, over a stationary discrete memoryless interference channel (DM-IC) without feedback has been extensively studied \cite{elgamalNetworkInformationTheory2011}.
A DM-IC is characterized as a conditional probability $P_{Y_{[K]} | X_{[K]}}$ from $\mathcal{X}_{[K]}$ to $\mathcal{Y}_{[K]}$, where $\mathcal{X}_i$ and $\mathcal{Y}_i$ are finite sets denoting an input alphabet at the $i$-th transmitter side and an output alphabet at the $i$-th receiver side, respectively.
This channel is stationary and memoryless in the sense that $n$-times use of the channel is given by $P_{Y_{[K]}^n | X_{[K]}^n} =P_{Y_{[K]}|X_{[K]}}^{\otimes n}$, i.e.,
\begin{equation}
    P_{Y_{[K]}^n | X_{[K]}^n} (y^n_{[K]} | x_{[K]}^n ) = \prod\limits_{j=1}^n P_{Y_{[K],j} | X_{[K],j}} (y_{[K],j}  | x_{[K],j}).
\end{equation}

For each $i \in [K]$, the $i$-th transmitter has its own message $M_i \in \mathcal{M}_i$ and transmits a codeword $X^n_i$ corresponding to $M_i$ over the channel over $n$ channel uses $P_{Y_{[K]}^n | X_{[K]}^n}$.
{\rev Here, the messages $M_{[K]}$ are assumed to be uniformly distributed.}
Then, the $i$-th receiver decodes the noisy channel output $Y_i^n$ into $\hat{M}_i$ to estimate the message $M_i$.

In this case, general communication strategy over a DM-IC $P_{Y_{[K]} | X_{[K]}}$ that transmitters and receivers can use is characterized by a $(n,R_{[K]})$-code, which is defined as a collection of
\begin{enumerate}
    \item Message sets $\mathcal{M}_i:=\left[ 2^{\lceil nR_i \rceil} \right]$ for all $i \in {\rev [K]}$,
    \item Encoding functions (encoders) $f_i :\mathcal{M}_i \rightarrow  \mathcal{X}_i^n$ for all $i \in {\rev [K]}$,
    \item Decoding functions (decoders) $g_i:\mathcal{Y}_i^n \rightarrow \mathcal{M}_i \cup \{e\}$ for all $i \in {\rev [K]}$, where $e$ denotes an error symbol.
\end{enumerate}
Note that for a given $(n,R_{[K]})$-code, we have $X^n_i = f_i(M_i)$ and $\hat{M}_i = g_i(Y^n_i)$ for all $i \in {\rev [K]}$.
We say that a rate tuple $R_{[K]}$ is achievable (for $P_{Y_{[K]} | X_{[K]}}$) if there exists a sequence of $(n,R_{[K]})$-codes such that the average error probability converges to zero, i.e.,
\begin{equation}\label{eq:avg_err_zero}
    \lim\limits_{n \rightarrow \infty} \mathrm{Pr}(M_{[K]} \neq \hat{M}_{[K]})=0.
\end{equation}
The capacity region $\mathcal{C}(P_{Y_{[K]} | X_{[K]}})$ of a DM-IC $P_{Y_{[K]} | X_{[K]}}$ is defined as the closure of the set of all achievable rate tuples, and the sum capacity $\mathcal{C}_s(P_{Y_{[K]} | X_{[K]}})$ is defined as
\begin{multline}
    \mathcal{C}_s(P_{Y_{[K]} | X_{[K]}})
    \\ := \max\left\{\sum_{i\in[K]}R_i : R_{[K]} \in \mathcal{C}(P_{Y_{[K]} | X_{[K]}}) \right\}.
\end{multline}

To date, a single-letter characterization of the capacity region and the sum capacity of a DM-IC is not known in general, even for $K=2$.
The best-known inner bound of the capacity region of a general DM-IC is the Han-Kobayashi inner bound \cite{hanNewAchievableRate1981,chongHanKobayashiRegion2008}, and this inner bound is tight for some class of DM-ICs.
{\rev
The following examples present channels whose sum capacities are known.
\begin{example}[Modulo-$2$ sum IC {\cite[Ex. 6.1]{elgamalNetworkInformationTheory2011}}]
    Consider two transmitter-receiver pairs, and all the input and output alphabets are binary, i.e., $\mathcal{X}_1 = \mathcal{X}_2 = \mathcal{Y}_1 = \mathcal{Y}_2 = \{0,1\}$.
    The interference between transmitted signals is modeled by 
    \begin{equation}
        Y_1 = Y_2 = X_1 \oplus X_2,
    \end{equation}
    or equivalently,
    \begin{equation}
        P_{Y_{[2]}|X_{[2]}} (y_{[2]} | x_{[2]}) = \delta_{y_1 , x_1 \oplus x_2} \delta_{y_2 , x_1 \oplus x_2},
    \end{equation}
    where $\delta$ denotes the Kronecker delta.
    For this channel, the sum capacity is upper bounded by $1$ since $Y_1 = Y_2$ and both $Y_1$ and $Y_2$ are binary.
    Also, the sum rate $1$ is achievable by a \emph{time sharing scheme}, i.e., the first transmitter keeps silent $(X_1= 0)$ for a fraction $\alpha \in [0,1]$ of the total channel uses, and the second transmitter keeps silent $(X_2 = 0)$ for the remaining $1-\alpha$ fraction.
    For any $\alpha$, either $X_1$ or $X_2$ can be perfectly communicated, and hence the sum rate $1$ is achievable.
\end{example}

\begin{example}[Parallel subchannels {\cite[Ex. 6.2]{elgamalNetworkInformationTheory2011}}]
    Consider a scenario with $K$ transmitter-receiver pairs.
    A channel $P_{Y_{[K]}|X_{[K]}}$ composed of parallel subchannels is described by 
    \begin{equation}
        P_{Y_{[K]}|X_{[K]}} (y_{[K]} | x_{[K]}) = \prod\limits_{i=1}^K P_{Y_i | X_i}(y_i | x_i ),
    \end{equation}
    i.e., for all $i \in [K]$, the received signal at the $i$-th receiver side $Y_i$ is not affected by any transmitted signal other than the transmitted signal from the $i$-th transmitter $X_i$.
    For this channel, the sum capacity is upper bounded by the sum of the point-to-point capacities of the subchannels $P_{Y_i | X_i}$ since it is the point-to-point capacity achievable by considering all the transmitters (resp. receivers) as a single joint transmitter (resp. receiver).
    Also, this upper bound is achievable by operating the point-to-point capacity-achieving communication strategy for each transmitter-receiver pair.
\end{example}

\begin{remark}\label{rmk:IC_to_MAC}
    A DM-IC $P_{Y_{[K]}|X_{[K]}}$ can be identified as a DM-MAC $P_{\tilde{Y}|X_{[K]}}$ by treating multiple receivers as a single receiver, i.e., $\tilde{\mathcal{Y}} = \mathcal{Y}_{[K]}$, and the single joint receiver observes the channel output $\tilde{Y}=Y_{[K]}$ and aims to infer all the messages $M_{[K]}$.
    With this identification, all results and proofs for DM-ICs presented in the rest of the paper apply directly to those for the corresponding DM-MACs.
    Therefore, we provide a detailed presentation of the results only for DM-ICs.
\end{remark}
}

\subsection{Communication over IC with transmitter cooperation}\label{sec:coop}
Now, we consider a scenario where transmitters cooperate using some pre-shared resources, which has been studied in \cite{quekQuantumSuperquantumEnhancements2017,YunRaiBae2020,  hawellekInterferenceChannelEntangled2025}.
With such cooperation, the transmitters can transmit correlated codewords, and the correlation between them can help mitigate interference and hence enlarge the capacity region.

Cooperation between the transmitters is characterized by a conditional probability that can be classified based on the type of pre-shared resource: classical, quantum, or super-quantum resource \cite{brunnerBellNonlocality2014}.
Cooperation that utilizes a classical, quantum, or super-quantum resource is termed classical, quantum, or no-signaling cooperation, respectively, and is defined as follows.

\begin{definition}
A conditional probability $P_{A_{[K]}|Q_{[K]}}$ is called
\begin{enumerate}
    \item a classical cooperation if there exists a finite set $\mathcal{V}$, a probability distribution $P_V$ supported on $\mathcal{V}$, and conditional distributions $P_{A_i|Q_i,V}$ for all $i \in [K]$ satisfying
    \begin{multline}\label{eq:C_coop}
        P_{A_{[K]}|Q_{[K]}}(a_{[K]} | q_{[K]})
        \\ = \sum\limits_{v \in \mathcal{V}} P_V(v) \prod\limits_{i=1}^K P_{A_i|Q_i,V}(a_i|q_i,v).
    \end{multline}

    \item a quantum cooperation if there exists a quantum state $\rho$ and measurements (positive operator-valued measures) $\{\Pi^i_{a_i|q_i}\}_{a_i \in \mathcal{A}_i}$ for all $q_i \in \mathcal{Q}_i$ and $i \in [K]$ satisfying
    \begin{equation}\label{eq:Q_coop}
        P_{A_{[K]}|Q_{[K]}}(a_{[K]} | q_{[K]}) = \mathrm{Tr}\left( \left(\otimes_{i=1}^K \Pi^i_{a_i|q_i}\right) \rho \right).
    \end{equation}

    \item a no-signaling cooperation if for all $i \in [K]$,
    \begin{equation}\label{eq:NS_coop}
        P_{A_i|Q_{[K]}}(a_i | q_{[K]})
        \\ = P_{A_i|Q_i}(a_i | q_i).
    \end{equation}
    Simply, for all $i \in [K]$, $(Q_1,\ldots,Q_{i-1},Q_{i+1},\ldots,Q_{K})$ and $A_i$ are conditionally independent given $Q_i$.
\end{enumerate}
We denote by $\mathcal{E}^{\mathrm{C}}, \mathcal{E}^{\mathrm{Q}}$, and $\mathcal{E}^{\mathrm{NS}}$ the sets of all classical, quantum, and no-signaling cooperations, respectively.
\end{definition}

Note that a quantum cooperation is not a classical cooperation only if a quantum state $\rho$ in \eqref{eq:Q_coop} is entangled.
Also, there is a strict hierarchy among the above classes of cooperation in the containment relation \cite{brunnerBellNonlocality2014}, i.e.,
\begin{equation}
    \mathcal{E}^\mathrm{C} \subsetneq \mathcal{E}^\mathrm{Q} \subsetneq \mathcal{E}^\mathrm{NS}. \label{eq:cont_hierarchy}
\end{equation}

In the communication scenario allowing cooperation between transmitters, a general communication strategy that transmitters and receivers can use is characterized by an $(n,R_{[K]},r)$-code.
The definition of a $(n,R_{[K]},r)$-code is the same as that of a $(n,R_{[K]})$-code described in Section~\ref{sec:DMIC}, except that the code contains cooperative encoders $P_{X^n_{[K]}|M_{[K]}}$ in $\mathcal{E}^r$ instead of encoding functions.
Accordingly, we say that a rate tuple $R_{[K]}$ is $r$-achievable if there exists a sequence of $(n,R_{[K]},r)$-codes satisfying \eqref{eq:avg_err_zero}, the $r$-capacity region $\mathcal{C}^r$ is the closure of the set of all $r$-achievable rate tuples, and the $r$-sum capacity $\mathcal{C}_s^r$ is the maximum of $\sum_{i \in [K]} R_i$ over all $R_{[K]} \in \mathcal{C}^r$.

In the case $K=2$, previous works \cite{quekQuantumSuperquantumEnhancements2017, hawellekInterferenceChannelEntangled2025, YunRaiBae2020} showed that for certain DM-ICs related to pseudo-telepathy games, utilizing quantum or no-signaling cooperation achieves a higher sum capacity than classical cooperation, i.e., $\mathcal{C}_s^\mathrm{C} < \mathcal{C}_s^\mathrm{Q}$ or $\mathcal{C}_s^\mathrm{C} < \mathcal{C}_s^\mathrm{NS}$. 

{\rev
\begin{remark}
    General no-signaling cooperation need not be physically implementable, but it serves as an outer benchmark for quantum cooperation. Since every quantum cooperation is no-signaling, the absence of a no-signaling advantage for a given channel rules out quantum advantage over classical cooperation.
    Thus, no-signaling cooperation provides a necessary-condition benchmark for quantum advantage.
    Conversely, a no-signaling advantage alone does not certify quantum advantage, since some no-signaling correlations may not be achievable by quantum resources.
\end{remark}
}

\subsection{Nonlocal games} \label{sec:game}
The usefulness of cooperation from a higher class in the hierarchy can be explained in a nonlocal game scenario \cite{brassardQuantumPseudoTelepathy2005a, cleveConsequencesLimitsNonlocal2004, brunnerBellNonlocality2014}.
A promise-free $K$-party game $G$ is defined by a tuple of finite sets $(\mathcal{Q}_{[K]},\mathcal{A}_{[K]}, \mathcal{W})$ where $\mathcal{W} \subseteq \mathcal{Q}_{[K]} \times \mathcal{A}_{[K]}$.
For each $i \in [K]$, $\mathcal{Q}_i$ and $\mathcal{A}_i$ denote the sets of questions and answers for the $i$-th player, respectively, and $\mathcal{W}$ denotes the winning condition.\footnote{A promise is a subset $P$ of the set of questions $\mathcal{Q}_{[K]}$. For a $K$-party game $G$ with promise $P$, a referee selects questions $Q_{[K]}$ only from $P$. In this paper, we only consider a promise-free game, i.e., $P = \mathcal{Q}_{[K]}$.}

\begin{figure}
    \centering
    \includegraphics[width=0.7\linewidth]{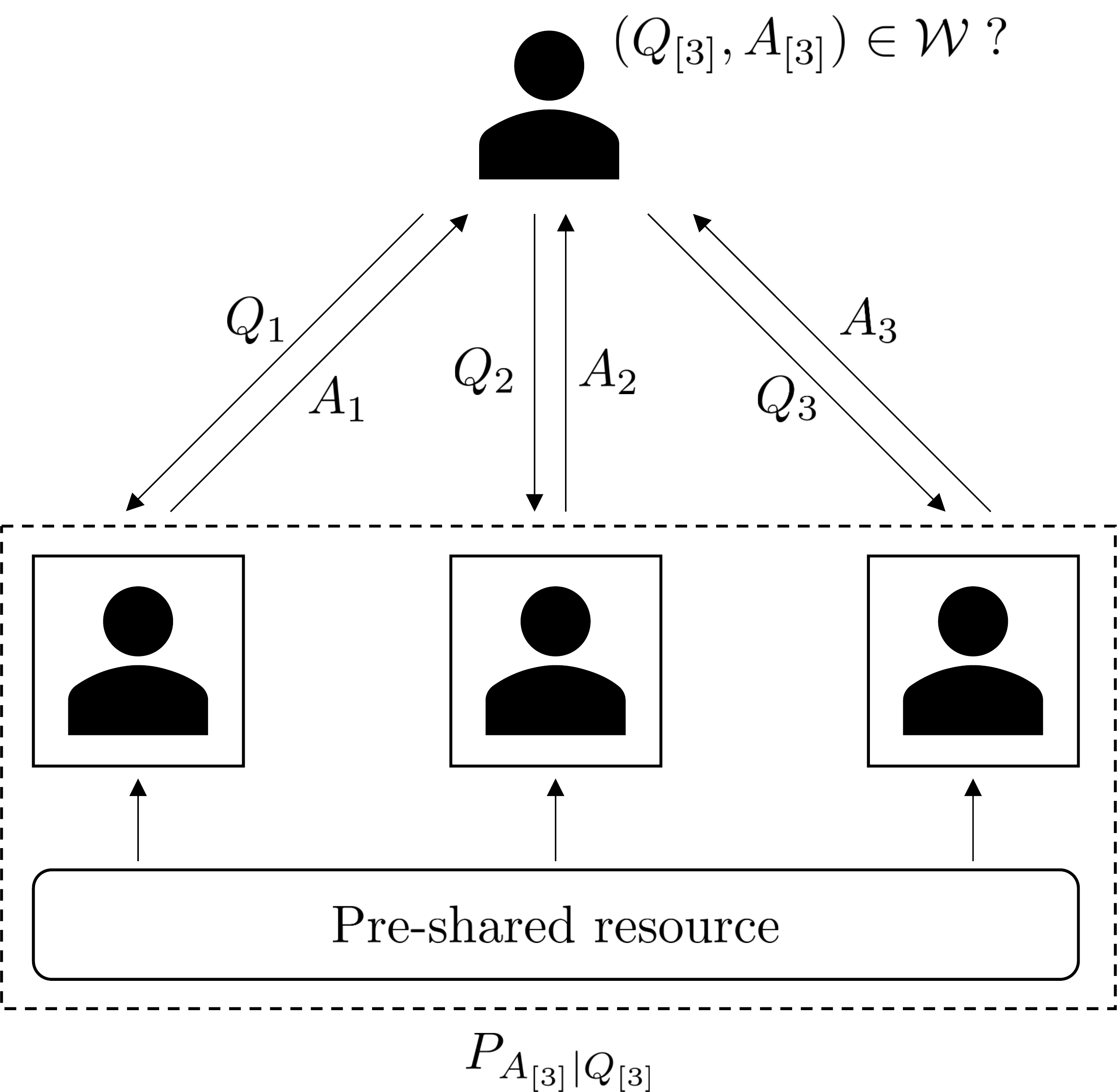}
    \caption{{\rev Nonlocal game scenario with $K=3$ players. A referee at the top gives questions $Q_1,Q_2,Q_3$ to each player, and players give answers $A_1,A_2,A_3$ based on a strategy $P_{A_{[3]}|Q_{[3]}}$ that utilizes a pre-shared resource.
    Players win the game if $(Q_{[3]},A_{[3]})$ satisfies the winning condition $\mathcal{W}$.
    }}
    \label{fig:game}
\end{figure}

In a scenario where $K$ players play a game $G$, {\rev as depicted in Figure~\ref{fig:game}}, all the players discuss and agree on a cooperation strategy $P_{A_{[K]}|Q_{[K]}}$ before the game starts.
After the game starts, each player enters an isolated room, i.e., they cannot communicate with each other from now on.
A referee then samples a uniformly random question tuple $Q_{[K]} \in \mathcal{Q}_{[K]}$ and gives $Q_i$ to the $i$-th player for all {\rev $i \in [K]$}.
Based on the predetermined cooperation strategy and the questions $Q_{[K]}$, the $i$-th player gives an answer $A_i$ to the referee.
The referee declares that the players win if $(Q_{[K]},A_{[K]}) \in \mathcal{W}$.

For a given game $G$, the usefulness of a cooperation $P_{A_{[K]}|Q_{[K]}}$ can be measured by the winning probability in the game.
Accordingly, the usefulness of the types of cooperations $\mathcal{E}^{r}$ can be measured by the maximum winning probability over $P_{A_{[K]}|Q_{[K]}} \in \mathcal{E}^{r}$ in the game, defined as
\begin{align}
    &p_w^r(G) := \max\limits_{P_{A_{[K]}|Q_{[K]}}\in \mathcal{E}^{r}} \mathrm{Pr}((Q_{[K]},A_{[K]}) \in \mathcal{W})
    \\& = \max\limits_{P_{A_{[K]}|Q_{[K]}}\in \mathcal{E}^{r}} \sum\limits_{(q_{[K]},a_{[K]}) \in \mathcal{W}} \frac{P_{A_{[K]}|Q_{[K]}} (a_{[K]} | q_{[K]} )}{|\mathcal{Q}_{[K]}|}.
\end{align}
If $p_w^r(G) = 1$ and $p_w^{\mathrm{C}}(G) < 1$ for some $r \in \{\mathrm{Q},\mathrm{NS}\}$, then we call a game $G$ an $r$-pseudo-telepathy game, and a cooperation $P_{A_{[K]}|Q_{[K]}}\in \mathcal{E}^{r}$ that achieves $p_w^r(G) = 1$ a winning strategy.

There are games that show strict gaps between the maximum winning probabilities that can be achieved by different types of cooperations.
One of the most representative examples is the Clauser–Horne–Shimony–Holt (CHSH) game \cite{clauserProposedExperimentTest1969}.
The CHSH game $G_{\mathrm{CHSH}}$ is a 2-party game such that 
\begin{equation}
    \forall i \in \{1,2\}, \; \mathcal{Q}_i = \mathcal{A}_i = \{0,1\},
\end{equation}
and
\begin{equation}
    \mathcal{W} = \{(q_{[2]},a_{[2]}): a_1 + a_2 \equiv q_1q_2 \; (\mathrm{mod}\; 2)\}.
\end{equation}
The CHSH game is a $\mathrm{NS}$-pseudo-telepathy game, and 
\begin{align}\label{eq:win_prob_hierarchy}
    p_w^{\mathrm{C}}(G_{\mathrm{CHSH}}) = 0.75 & < p_w^{\mathrm{Q}}(G_{\mathrm{CHSH}}) \approx 0.85
    \\ &< p_w^{\mathrm{NS}}(G_{\mathrm{CHSH}}) = 1.
\end{align}
Therefore, the CHSH game translates the strict hierarchy in the maximum winning probabilities to the strict hierarchy in the containment relation in \eqref{eq:cont_hierarchy}.
{\rev
Here, $p_w^{\mathrm{Q}}$ can be achieved by sharing a maximally entangled state between the players, and $p_w^{\mathrm{NS}}$ is achieved by a cooperation $P_{A_{[2]}|Q_{[2]}}$ called a Popescu-Rohrlich (PR) box \cite{popescuQuantumNonlocalityAxiom1994}.
}

One example of a $\mathrm{Q}$-pseudo-telepathy game is the magic square game $G_{\mathrm{MS}}$ \cite{aravindBellsTheoremInequalities2002}.
It is a two-party game that consists of
\begin{equation}
    \forall i \in \{1,2\},\; \mathcal{Q}_i = \{1,2,3\}, \; \mathcal{A}_i = \{0,1\}^3,
\end{equation}
and
\begin{equation}
    \!\!\!\!\!\! \mathcal{W} = \left\{(q_{[2]},a_{[2]}): \begin{matrix}
        \sum_{j \in [3]} a_{1,j} \equiv 0 \; (\mathrm{mod}\;2),
        \\ \sum_{j \in [3]} a_{2,j} \equiv 1 \; (\mathrm{mod}\;2),
        \\ a_{1,q_2} = a_{2,q_1}
    \end{matrix} \right\}.
\end{equation}
The magic square game also demonstrates a separation between winning probabilities,
\begin{equation}
    p_w^{\mathrm{C}}(G_{\mathrm{MS}}) = 8/9  < p_w^{\mathrm{Q}}(G_{\mathrm{MS}}) = 1.
\end{equation}
{\rev
The quantum maximum winning probability $p_w^{\mathrm{Q}}$ can be achieved by sharing a maximally entangled state between the players \cite{brassardQuantumPseudoTelepathy2005a}.
}

Beyond two-party $\mathrm{Q}$-pseudo-telepathy games, there are $K$-party $\mathrm{Q}$-pseudo-telepathy games for all $K\geq 3$, known as the (promise-free version of) $K$-party parity ($K$-PP) game \cite{brassardQuantumPseudoTelepathy2005a,seshadriSeparationCorrelationAssistedSum2023}.
The $K$-PP game $G_{K\mathrm{-PP}}$ consists of
\begin{equation}
    \forall i \in [K],\; \mathcal{Q}_i = \mathcal{A}_i = \{0,1\},
\end{equation}
and
\begin{align}
    &\mathcal{W} = \left\{(q_{[K]},a_{[K]}) : \sum\limits_{i \in [K]} q_i \equiv 1 \; (\mathrm{mod}\; 2) \right\}
    \\& \quad \cup
    \left\{
        (q_{[K]},a_{[K]}) : \begin{matrix}
            \sum\limits_{i \in [K]} q_i \equiv 0 \; (\mathrm{mod}\; 2),
        \\ \sum\limits_{i \in [K]} a_i \equiv \frac{1}{2}\sum\limits_{i \in [K]} q_i \; (\mathrm{mod}\; 2) 
        \end{matrix}
    \right\}.
\end{align}
For $G_{K\mathrm{-PP}}$ and $K \geq 3$, we have
\begin{align}
    p_w^\mathrm{C} (G_{K\mathrm{-PP}}) &= \frac{3}{4} + 2^{-(\lceil K/2 \rceil + 1)}
    \\& < p_w^\mathrm{Q} (G_{K\mathrm{-PP}}) = 1.
\end{align}
{\rev
Again, $p_w^\mathrm{Q}$ can be achieved by sharing a maximally entangled state among the players \cite{brassardMultipartyPseudoTelepathy2003}.
}

\section{Main results}

Our main result shows the enhancement of sum capacity according to the hierarchy of cooperation, i.e., $\mathcal{C}_s^{\mathrm{C}} < \mathcal{C}_s^{\mathrm{Q}}$ or $\mathcal{C}_s^{\mathrm{C}} < \mathcal{C}_s^{\mathrm{NS}}$, for a certain class of DM-ICs and DM-MACs.
The proposed class includes all the channels previously used to demonstrate such separation in previous works \cite{leditzkyPlayingGamesMultiple2020, seshadriSeparationCorrelationAssistedSum2023, peregMultipleAccessChannelEntangled2025, hawellekInterferenceChannelEntangled2025}, as well as many other channels beyond the channels considered in \cite{YunRaiBae2020, leditzkyPlayingGamesMultiple2020, hawellekInterferenceChannelEntangled2025, peregMultipleAccessChannelEntangled2025, seshadriSeparationCorrelationAssistedSum2023, 
quekQuantumSuperquantumEnhancements2017, yunNonLocalQuantumAdvantages2023}.

\subsection{DM-ICs from nonlocal games}

Inspired by $K$-party games and previous works \cite{YunRaiBae2020, leditzkyPlayingGamesMultiple2020, hawellekInterferenceChannelEntangled2025, peregMultipleAccessChannelEntangled2025, seshadriSeparationCorrelationAssistedSum2023, quekQuantumSuperquantumEnhancements2017, yunNonLocalQuantumAdvantages2023}, we propose a class of DM-ICs for which quantum or no-signaling cooperation can achieve a strictly larger sum capacity than classical cooperation.
{\rev As in previous works, we consider a channel whose input at transmitter
$i$ is a pair $x_i=(q_i,a_i)$ consisting of a question and an answer for a given nonlocal game $G$.
Here, $q_i$ can be interpreted as the message-bearing component of the transmitted signal, while $a_i$ can be viewed as a finite-alphabet coordination variable, analogous to a precoding or beamforming choice.
Under this interpretation, the winning condition $x_{[K]} \in \mathcal{W}$ acts as an interference-alignment condition: when the transmitters choose answers that satisfy the winning condition, the residual cross-link interference is aligned or suppressed at the receivers.

Previous works often considered channels in which, whenever the inputs satisfy the winning condition of $G$, each receiver obtains the corresponding question symbol noiselessly, or through simple random erasure or uniform-random-noise models.
Moving beyond these specific models, we formalize a broader notion of weaker interference in the winning mode through two features: reduced conditional output uncertainty and removal of cross-link dependence.

First, conditional output uncertainty is measured by the conditional entropy $H(Y_{[K]}|X_{[K]}=x_{[K]})$.
We require the winning mode to have strictly smaller conditional output uncertainty than every non-winning mode.
Thus, when $x_{[K]}\in\mathcal W$, the channel induces less uncertainty than when $x_{[K]}\notin\mathcal W$.

Second, to model the removal of cross-link interference in the winning
mode, we require the channel to decompose into parallel subchannels:
for every $(q_{[K]},a_{[K]})\in\mathcal W$,
\begin{equation}
    P^G_{Y_{[K]}|Q_{[K]},A_{[K]}}(y_{[K]}|q_{[K]},a_{[K]}) = \prod_{i\in[K]}P^{G}_{Y_i|Q_i}(y_i|q_i).
\end{equation}
Physically, this means that, once the winning condition is satisfied, the question component of transmitter $j$ does not affect the signal observed at receiver $i$ for any $j\neq i$.
Hence receiver $i$ observes only a noisy version of its intended question $Q_i$.
In a non-winning mode, this alignment may fail, and the unresolved components generated by other transmitters can be viewed as residual interference, or as additional effective noise under a treat-interference-as-noise interpretation.

To cover subchannels beyond noiseless, erasure, or uniform random noise models, we allow each winning-state subchannel $P^{G}_{Y_i|Q_i}$ to be weakly symmetric\cite[Sec.~7.2]{coverElementsInformationTheory2012}.
We further impose that the uniform distribution on $\mathcal Q_i$ is its unique capacity-achieving input distribution.
This uniqueness condition will be used in the converse proof to show that attaining the sum capacity upper bound would require a perfect classical winning strategy for the underlying game.
Physically motivated examples include uniform random noise channels and cyclic nearest-neighbor confusion channels.
The formal description is as follows.

\begin{definition}
    A conditional probability $P_{Y|Q}$ is weakly symmetric if 
    \begin{enumerate}
        \item $\sum_{q \in \mathcal{Q}} P_{Y|Q}(y|q) $ is constant for all $y \in \mathcal{Y}$,
        \item For any $q,q' \in \mathcal{Q}$, there exists a permutation $\sigma$ on $\mathcal{Y}$ such that $P_{Y|Q}(\cdot|q') = P_{Y|Q}(\sigma(\cdot)|q)$.
    \end{enumerate}
\end{definition}

A useful property of a weakly symmetric channel is that its capacity has a closed-form expression as in the following lemma.

\begin{lemma}[{\cite[Lem.~4.16]{alajajiIntroductionSingleUserInformation2018}}]\label{lem:weak_sym_ch}
    If $P_{Y|Q}$ is weakly symmetric, then $H(Y|Q=q)$ is constant over $q \in \mathcal{Q}$ and the point-to-point channel capacity of $P_{Y|Q}$ is given by
    \begin{equation}
        \mathcal{C}(P_{Y|Q}) := \max\limits_{P_Q} I(Q;Y) = \log |\mathcal{Y}| - H(Y|Q=q),
    \end{equation}
    where $I(Q;Y)$ denotes the mutual information between $Q$ and $Y$.
    Moreover, the maximum is achieved when $P_Q$ is the uniform distribution on $\mathcal{Q}$, and $P_Y$ becomes the uniform distribution on $\mathcal{Y}$ when $P_Q$ is the uniform distribution on $\mathcal{Q}$.
\end{lemma}
The following are examples of weakly symmetric channels.

\begin{example}[Uniform random noise channel]\label{ex:unif_ch}
    Let $\mathcal Q=\mathcal Y=[m]$. For $\eta\in[0,1]$, define a channel
    $P_{Y|Q}$ by
    \begin{equation}
        P_{Y|Q}(y|q) = \eta\delta_{y,q} + \frac{1-\eta}{m}.
    \end{equation}
    Equivalently, with probability $\eta$, the output equals the input     symbol $q$, while with probability $1-\eta$, the output is replaced by a uniformly random symbol in $[m]$.
    
    This channel is weakly symmetric.
    In detail, for every $y\in[m]$,
    \begin{equation}
        \sum_{q\in[m]}P_{Y|Q}(y|q) = \eta+\sum_{q\in[m]}\frac{1-\eta}{m} = 1,
    \end{equation}
    which is independent of $y$.
    Also, for any $q,q' \in [m]$, there exists a permutation $\sigma$ on $[m]$ such that $q' = \sigma(q)$.
    For any such permutation, we have
    \begin{equation}
        P_{Y|Q}(\cdot | \sigma(q)) = P_{Y|Q}(\sigma^{-1}(\cdot) | q).
    \end{equation}
    Hence, $P_{Y|Q}$ is weakly symmetric.
    
    By Lemma~\ref{lem:weak_sym_ch}, the conditional entropy is independent of $q$ and is given by
    \begin{equation}
        H(Y|Q=q) = H_u(\eta,m),
    \end{equation}
    where
    \begin{align}
        H_u(\eta,m)  &:=
        -\left(\eta+\frac{1-\eta}{m}\right)\log\left(\eta+\frac{1-\eta}{m}\right) \label{eq:H_u_def}
        \\& \quad -(m-1)\frac{1-\eta}{m}\log\left(\frac{1-\eta}{m}\right).  \nonumber
    \end{align}
    Also, the capacity is given by
    \begin{equation}
        C(P_{Y|Q}) = \log m-H_u(\eta,m).
    \end{equation}
    Furthermore, if $\eta>0$, then the uniform input distribution is the
    unique capacity-achieving input distribution because, when $P_{Y|Q}$ is identified as a transition probability matrix, it is invertible.
\end{example}

\begin{example}[Cyclic nearest-neighbor confusion channel]\label{ex:nearest_neighbor}
    Let $\mathcal Q=\mathcal Y=\mathbb Z_m:= \{0,1,\dots,m-1\}$, where $m\ge 3$.
    For $\epsilon\in [1/2,1]$, define a channel $P_{Y|Q}$ by
    \begin{equation}
        P_{Y|Q}(y|q) = \begin{cases}
            \epsilon & \text{ if }y=q\\
            (1-\epsilon)/2 & \text{ if } y=q \pm 1 \pmod m\\
            0 & \text{ otherwise }
        \end{cases},
    \end{equation}
    which can be conceptually visualized as in Figure~\ref{fig:nearest}.
    This channel can be interpreted as a simplified hard-decision model for an $m$-phase-shift keying (PSK) constellation, where the receiver detects the transmitted symbol correctly with probability $\epsilon$ and otherwise confuses it with one of its two adjacent symbols.
    \begin{figure}
        \centering
        \includegraphics[width=0.35\linewidth]{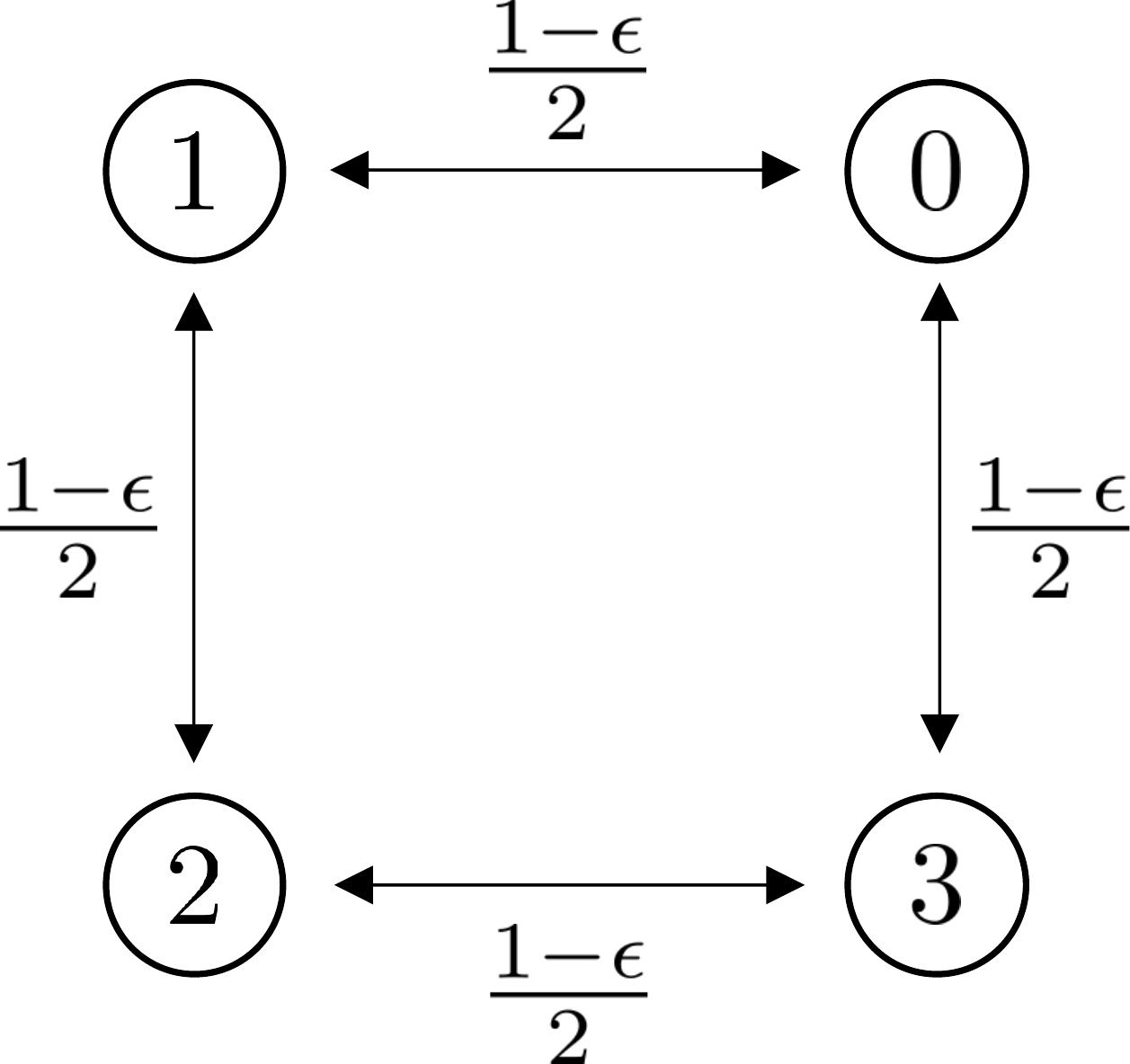}
        \caption{{\rev A cyclic nearest-neighbor confusion channel in Example~\ref{ex:nearest_neighbor} with $m=4$.
        Each symbol can be confused to one of two nearest-neighboring symbols with probability $(1-\epsilon)/2$.
        }}
        \label{fig:nearest}
    \end{figure}
    
    The channel is weakly symmetric.
    In detail, for each $y\in\mathbb Z_m$, we have
    \begin{equation}
        \sum_{q\in\mathbb Z_m} P_{Y|Q}(y|q) = \epsilon+\frac{1-\epsilon}{2}+\frac{1-\epsilon}{2} = 1,
    \end{equation}
    which is independent of $y$.
    Also, for any $q,q' \in \mathbb{Z}_m$, there exists a cyclic shift permutation $\sigma$ on $\mathbb{Z}_m$ such that $\sigma(q) = q'$.
    Then, we have 
    \begin{equation}
        P_{Y|Q}(\cdot|q') = P_{Y|Q}(\cdot|\sigma(q)) = P_{Y|Q}(\sigma^{-1}(\cdot)|q).
    \end{equation}
    Therefore, $P_{Y|Q}$ is weakly symmetric.
    
    By Lemma~\ref{lem:weak_sym_ch}, the conditional entropy is independent of $q$ and is given by
    \begin{equation}
        H(Y|Q=q) = H_2(\epsilon)+1-\epsilon,
    \end{equation}
    where $H_2$ is the binary entropy function,
    \begin{equation}
        H_2(\epsilon) := -\epsilon \log \epsilon - (1-\epsilon) \log (1-\epsilon).
    \end{equation}
    Also, the capacity of this channel is given by
    \begin{equation}
        \mathcal{C}(P_{Y|Q}) = \log m-H_2(\epsilon)-(1-\epsilon).
    \end{equation}
    Furthermore, if $\epsilon > 1/2$, then the uniform input distribution is the unique capacity-achieving input distribution because by identifying $P_{Y|Q}$ as a transition matrix, it is an invertible matrix.
\end{example}

We now formally define the proposed class of DM-ICs.
\begin{definition}\label{def:game_ch}
    Let $G=(\mathcal{Q}_{[K]},\mathcal{A}_{[K]},\mathcal{W})$ be a $K$-party game.
    A DM-IC $P_{Y_{[K]}|X_{[K]}}^{G}$ is called a $G$-game channel if $\mathcal{X}_i = \mathcal{Q}_i\times \mathcal{A}_i$, $\mathcal{Y}_{i}$ is finite for every $i \in [K]$, and the following conditions hold:
    \begin{enumerate}
        \item \emph{Aligned winning mode}: 
        For every $(q_{[K]},a_{[K]}) \in \mathcal{W}$,
        \begin{multline}\label{eq:parallel_ch}
            P^{G}_{Y_{[K]}|Q_{[K]},A_{[K]}}(y_{[K]}|q_{[K]},a_{[K]})
            \\ = \prod\limits_{i\in[K]} P^{G}_{Y_i|Q_i}(y_i|q_i).
        \end{multline}
    
        \item \emph{Weakly symmetric subchannels with unique capacity-achieving input distribution}: For every $i \in [K]$, the local winning-state subchannel $P^{G}_{Y_i|Q_i}$ is weakly symmetric, and its capacity-achieving input distribution is unique.
        Let
        \begin{equation}\label{eq:win_cond_ent}
            h_i^w := H(Y_i|Q_i=q_i),
        \end{equation}
        which is independent of $q_i$ by weak symmetry.

        \item \emph{Strictly larger conditional output entropy outside the winning mode:}
        For every $x_{[K]} \in \mathcal{W}$ and $\tilde{x}_{[K]}\notin\mathcal{W}$,
        \begin{align}\label{eq:sum_cond_ent_less_noisy}
            H(Y_{[K]}|X_{[K]}=\tilde{x}_{[K]}) &> H(Y_{[K]}|X_{[K]}=x_{[K]})
            \\& = \sum_{i\in[K]}h_i^w .
        \end{align}
    \end{enumerate}
\end{definition}

\subsection{Examples}\label{sec:ex_G_ch}
There exist many $G$-game channels for a fixed game $G$.
We provide two representative examples based on uniform random noise channel in Example~\ref{ex:unif_ch} and cyclic nearest-neighbor confusion channel in Example~\ref{ex:nearest_neighbor}, as depicted in Figure~\ref{fig:G_ch_ex}.

For both examples below, fix a $K$-party game $G=(\mathcal{Q}_{[K]},\mathcal{A}_{[K]},\mathcal{W})$.
Let $\mathcal{X}_i = \mathcal{Q}_i \times \mathcal{A}_i$.
The channel first maps $x_{[K]} = (q_{[K]},a_{[K]})$ to an effective question tuple 
\begin{equation}
    \tilde{q}_{[K]} = (\tau_1(x_{[K]}),\dots,\tau_K(x_{[K]})),
\end{equation}
where each $\tau_i : \mathcal{X}_{[K]} \rightarrow \mathcal{Q}_i$ satisfies 
\begin{equation}\label{eq:tau_cond}
    \tau_i (x_{[K]}) = q_i, \quad \forall i \in [K],\; x_{[K]} \in \mathcal{W}.
\end{equation}
Thus, in the winning mode, the effective input observed by the $i$-th local subchannel coincides with the intended question $q_i$, whereas in the non-winning mode $\tau_i(x_{[K]})$ may include residual cross-link interference.

Then, we define
\begin{multline}
    P^G_{Y_{[K]}|X_{[K]}}(y_{[K]}|x_{[K]})
    \\ = \begin{cases}
        \prod_{i\in[K]} P^{w,i}_{Y_i|\tilde{Q}_i}(y_i|\tau_i(x_{[K]})) & \text{ if } x_{[K]} \in \mathcal{W} \\
        \prod_{i\in[K]} P^{l,i}_{Y_i|\tilde{Q}_i}(y_i|\tau_i(x_{[K]})) & \text{ if } x_{[K]} \notin \mathcal{W}
    \end{cases}.
\end{multline}
From \eqref{eq:tau_cond}, we have 
\begin{equation}
    P_{Y_i | \tilde{Q}_i}^{w,i} (y_i | \tau_i (x_{[K]})) = P_{Y_i | Q_i}^{w,i} (y_i | q_i),\quad \forall i \in [K],
\end{equation}
in the winning mode.
Whenever $P_{Y_i | \tilde{Q}_i}$ induces less output conditional entropy in the winning mode, this channel forms a $G$-game channel.
The detailed constructions are as follows.

\begin{figure}
    \centering
    \includegraphics[width=0.7\linewidth]{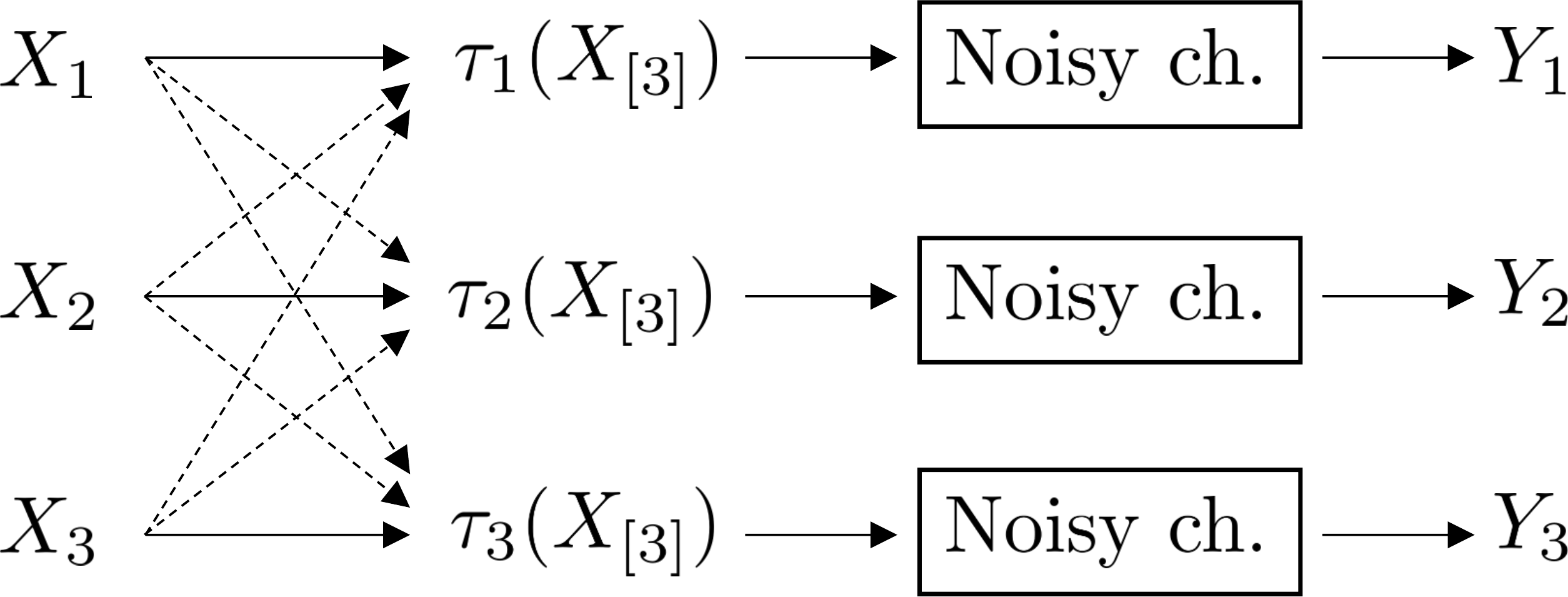}
    \caption{{\rev An example of a $G$-game channel described in Section~\ref{sec:ex_G_ch} with $K=3$.
    If $X_{[3]}$ satisfies the winning condition, then $\tau_i(X_{[3]}) = Q_i$, and the channel between $\tau_i (X_{[3]})$ and $Y_i$ has less conditional output uncertainty.
    }}
    \label{fig:G_ch_ex}
\end{figure}

\subsubsection{$G$-game channels with uniform random noise}\label{ex:game_ch_ex_unif}

Suppose $\mathcal{Q}_i = \mathcal{Y}_i = [m]$, $m \geq 2$, for every $i \in [K]$.
For $\eta_i\in[0,1]$, let the uniform random noise channel
\begin{equation}
    P^{\eta_i}_{Y_i|\tilde{Q}_i}(y_i|q_i) = \eta_i\delta_{y_i,q_i} + \frac{1-\eta_i}{m}, \quad q_i,y_i\in[m].
\end{equation}
Choose parameters $0\leq \eta_i^l < \eta_i^w \leq 1$.
Then, we set $P_{Y_i | \tilde{Q}_i}^{w,i} = P_{Y_i|\tilde{Q}_i}^{\eta_i^w}$ and $P_{Y_i | \tilde{Q}_i}^{l,i} = P_{Y_i|\tilde{Q}_i}^{\eta_i^l}$.

By Example~\ref{ex:unif_ch}, each $P_{Y_i | Q_i}^{\eta_i^w}$ is a weakly symmetric subchannel and has the uniform input distribution as its unique capacity-achieving distribution because $\eta_i^w>0$.
Also, the output conditional entropy is given by 
\begin{multline}
    H(Y_{[K]}|X_{[K]}=x_{[K]})
    \\ = \begin{cases}
        \sum_{i\in[K]}H_u(\eta_i^w,m) & \text{ if } x_{[K]} \in \mathcal{W} \\
        \sum_{i\in[K]}H_u(\eta_i^l,m) & \text{ if } x_{[K]} \notin \mathcal{W}
    \end{cases}.
\end{multline}
where $H_u(\eta,m)$ is defined in \eqref{eq:H_u_def}.
Since $H_u(\eta,m)$ is strictly decreasing in $\eta$ and $\eta_i^l<\eta_i^w$ for every $i\in[K]$, the strict conditional output uncertainty condition \eqref{eq:sum_cond_ent_less_noisy} holds.
Therefore, this channel is a $G$-game channel.

The special case $\eta_i^w=1$ and $\eta_i^l=0$ for all $i \in [K]$ recovers the noiseless-winning/uniform-losing constructions used in several previous works \cite{leditzkyPlayingGamesMultiple2020, seshadriSeparationCorrelationAssistedSum2023, peregMultipleAccessChannelEntangled2025, hawellekInterferenceChannelEntangled2025}.
Allowing $0\leq \eta_i^l < \eta_i^w \leq 1$ also permits noisy winning modes and non-winning modes with residual interference through the maps $\tau_i$.
Hence, this example extends the previously considered constructions based on uniform random noise.

\subsubsection{$G$-game channels with cyclic nearest-neighbor confusion}\label{ex:game_ch_nearest}
We next instantiate the same construction with cyclic nearest-neighbor confusion channel.
Suppose $\mathcal Q_i=\mathcal Y_i=\mathbb Z_m, \quad m\ge 3$, for every $i \in [K]$.
For $\epsilon_i \in [1/2,1]$, define the cyclic nearest-neighbor confusion channel in Example~\ref{ex:nearest_neighbor},
\begin{multline}
    P_{Y_i|\tilde{Q}_i}(y_i|q_i)
    \\ = \begin{cases}
        \epsilon_i & \text{ if }y_i=q_i\\
        (1-\epsilon_i)/2 & \text{ if } y_i=q_i \pm 1 \pmod m\\
        0 & \text{ otherwise }.
    \end{cases}
\end{multline}
Choose $1/2 \leq \epsilon_i^l < \epsilon_i^w \leq 1$, and set
\begin{equation}
    P^{w,i}_{Y_i|\tilde{Q}_i} = P^{\epsilon_i^w}_{Y_i|\tilde{Q}_i}, \; P^{l,i}_{Y_i|\tilde{Q}_i} = P^{\epsilon_i^l}_{Y_i|\tilde{Q}_i}.
\end{equation}

As shown in Example~\ref{ex:nearest_neighbor}, the winning-state subchannel $P^{\epsilon_i^w}_{Y_i|Q_i}$ is weakly symmetric and has the uniform input as its unique capacity-achieving distribution because $\epsilon_i^w>1/2$.
Also, the output conditional entropy is given by 
\begin{multline}
    H(Y_{[K]}|X_{[K]}=x_{[K]})
    \\ = \begin{cases}
        \sum_{i\in[K]}(H_2(\epsilon_i^w)+(1-\epsilon_i^w)) & \text{ if } x_{[K]} \in \mathcal{W} \\
        \sum_{i\in[K]}(H_2(\epsilon_i^l)+(1-\epsilon_i^l)) & \text{ if } x_{[K]} \notin \mathcal{W}
    \end{cases}.
\end{multline}
Since $\epsilon\mapsto H_2(\epsilon)+(1-\epsilon)$ is strictly decreasing in $\epsilon \in [1/2,1]$ and $\epsilon_i^l<\epsilon_i^w$ for every $i\in[K]$, the strict conditional output uncertainty condition \eqref{eq:sum_cond_ent_less_noisy} follows.
Thus, this channel is also a $G$-game channel.
}

\subsection{Enhancing sum capacity}

We show that quantum or no-signaling cooperation can achieve a higher sum capacity than classical cooperation for a $G$-game channel, when $G$ is a $\mathrm{Q}$- or $\mathrm{NS}$-pseudo-telepathy game.

\begin{theorem}\label{thm:main}
    Let $G$ be a $K$-party game and $P_{Y_{[K]}|X_{[K]}}^{G}$ be a $G$-game channel.
    For $r \in \{\mathrm{Q},\mathrm{NS}\}$, if $G$ is an $r$-pseudo-telepathy game, then
    \begin{equation}
        \mathcal{C}_s^\mathrm{C}(P_{Y_{[K]}|X_{[K]}}^{G}) < \mathcal{C}_s^r(P_{Y_{[K]}|X_{[K]}}^{G}).
    \end{equation}
\end{theorem}

As introduced in Section~\ref{sec:game}, there are $Q$-pseudo-telepathy $K$-party games for all $K \geq 2$ (the magic square game for $K=2$, and the $K$-party parity game for $K \geq 3$).
Thus, Theorem~\ref{thm:main} implies that, for all $K\geq 2$, there is a DM-IC $P_{Y_{[K]}|X_{[K]}}$ which shows an advantage of quantum cooperation over classical cooperation.

{\rev
Intuitively, Theorem~\ref{thm:main} follows from the facts that a $G$-game channel induces \emph{weaker interference} when the channel inputs satisfy the winning condition of a pseudo-telepathy game $G$, and quantum or no-signaling cooperation can perfectly win the game while no classical cooperation can.
}
In the remainder of this subsection, we {\rev formally} prove Theorem~\ref{thm:main}.
We begin by deriving an upper bound on $\mathcal{C}_s^r$ of a $G$-game channel (the converse part).
This upper bound is strict if $r= \mathrm{C}$ and $G$ is a pseudo-telepathy game.
Next, we prove that this upper bound is $r$-achievable when $G$ is an $r$-pseudo-telepathy game by exploiting the cooperation corresponding to a winning strategy for $G$ (the achievability part).
Combining the converse and achievability proofs, we establish Theorem~\ref{thm:main}.

\subsubsection{Converse}

We propose an upper bound on the $r$-sum capacity of a game channel as follows.
\begin{proposition}\label{prop:converse}
    For all $G$-game channels $P_{Y_{[K]}|X_{[K]}}^{G}$ and $r \in \{\mathrm{C,Q,NS}\}$, we have
    \begin{equation}
        \mathcal{C}_s^{r}\left( P_{Y_{[K]}|X_{[K]}}^{G} \right) \leq \sum\limits_{i \in [K]} (\log|\mathcal{Y}_i| - h_i^w), \label{eq:conv_gen}
    \end{equation}
    {\rev where $h_i^w$ is defined in \eqref{eq:win_cond_ent}.}
    Moreover, the above inequality is strict when $r = \mathrm{C}$ and $G$ is a pseudo-telepathy game.
\end{proposition}

The proof of the above proposition utilizes the fact that classical cooperation does not enlarge the capacity region.
\begin{lemma}\label{lem:C_eq_C_s}
    For all DM-ICs $P_{Y_{[K]}|X_{[K]}}$, we have that $\mathcal{C}(P_{Y_{[K]}|X_{[K]}}) = \mathcal{C}^\mathrm{C}(P_{Y_{[K]}|X_{[K]}})$.
\end{lemma}
\begin{IEEEproof}
    Let $P_{X_{[K]}^n|M_{[K]}} \in \mathcal{E}^{\mathrm{C}}$ be a cooperative encoder that is utilized in a $(n,R_{[K]},\mathrm{C})$-code.
    Since the average error probability $\mathrm{Pr}(M_{[K]} \neq \hat{M}_{[K]})$ is linear in $P_{X_{[K]}^n|M_{[K]}}$, one of the extreme points of $\mathcal{E}^{\mathrm{C}}$ achieves the minimum error probability.
    An extreme point of $\mathcal{E}^{\mathrm{C}}$ is equivalent to the collection of deterministic encoding functions $f_1,\ldots,f_K$ that can be performed by the transmitters without cooperation.
    Hence, there exists an $(n,R_{[K]})$-code whose average error probability is not greater than that of the given $(n,R_{[K]},\mathrm{C})$-code.
\end{IEEEproof}

Now, we prove Proposition~\ref{prop:converse}.
\begin{IEEEproof}[Proof of Proposition~\ref{prop:converse}]
    For any sequence of $(n,R_{[K]},r)$-codes or $(n,R_{[K]})$-codes that achieves $R_{[K]}$, we have
    \begin{align}
        n & \sum\limits_{i \in [K]} R_i = H(M_{[K]})
        \\&\leq I(M_{[K]} ; Y_{[K]}^n) + n \epsilon_n \label{eq:conv_fano}
        \\& \leq I(X_{[K]}^n;Y_{[K]}^n) + n \epsilon_n \label{eq:conv_DPI}
        \\& \leq \sum\limits_{j\in[n]}I(X_{[K],j};Y_{[K],j}) + n \epsilon_n \label{eq:conv_MI_subadd}
        \\& = \sum\limits_{j\in[n]} (H(Y_{[K],j}) - H(Y_{[K],j}|X_{[K],j})) + n\epsilon_n
        \\& \leq n \sum\limits_{i\in[K]} \log |\mathcal{Y}_i| - \sum\limits_{j=1}^n H(Y_{[K],j}|X_{[K],j}) + n \epsilon_n,\label{eq:conv_cond_ent}
    \end{align}
    where \eqref{eq:conv_fano} follows from Fano's inequality with $\epsilon_n \rightarrow 0$ as $n \rightarrow \infty$.
    Also, the data-processing inequality implies \eqref{eq:conv_DPI}, and \eqref{eq:conv_MI_subadd} holds since the channel is memoryless.
    The conditional entropy terms in \eqref{eq:conv_cond_ent} can be bounded as follows:
    \begin{align}
        \nonumber & H (Y_{[K],j}|X_{[K],j})
        \\& = \sum\limits_{x_{[K],j} \in \mathcal{W}} P_{X_{[K],j}}(x_{[K],j})H(Y_{[K],j}|X_{[K],j} = x_{[K],j})
        \\& \nonumber \quad + \sum\limits_{x_{[K],j} \not\in \mathcal{W}} P_{X_{[K],j}}(x_{[K],j})H(Y_{[K],j}|X_{[K],j} = x_{[K],j})
        \\& \geq P_{X_{[K],j}}(\mathcal{W})\sum\limits_{i \in [K]} h_i^w \label{eq:conv_C_strict}
        + (1-P_{X_{[K],j}}(\mathcal{W}))\sum\limits_{i \in [K]} h_i^w
        \\& = \sum\limits_{i \in [K]} h_i^w,
    \end{align}
    where \eqref{eq:conv_C_strict} follows from \eqref{eq:win_cond_ent} and \eqref{eq:sum_cond_ent_less_noisy}.
    Thus, we have \eqref{eq:conv_gen}.

    Now, suppose $r=\mathrm{C}$ and $G$ is a pseudo-telepathy game, i.e., there is no perfect classical winning strategy.
    By Lemma~\ref{lem:C_eq_C_s}, it is enough to consider a scenario that the transmitters do not cooperate.
    {\rev Hence, for every channel use $j$, the input distribution has product form $P_{X_{[K],j}}=\prod_{i\in[K]}P_{X_{i,j}}$.
    Let
    \begin{equation}
        U:=\sum_{i\in[K]}(\log|\mathcal Y_i|-h_i^w).
    \end{equation}
    Fix an arbitrary channel use $j$ and suppose, to the contrary, that $I(X_{[K],j};Y_{[K],j})=U$.
    Then, from \eqref{eq:conv_cond_ent} and \eqref{eq:conv_C_strict}, $Y_{i,j}$ must be uniformly distributed over $\mathcal{Y}_i$ for all $i$ and $P_{X_{[K],j}} (\mathcal{W}) = 1$ must hold.
    Here, $P_{X_{[K],j}}(\mathcal W)=1$ implies that the channel operates only in the aligned winning mode.
    Also, because $Y_{i,j}$ is uniform and $I(X_{[K],j};Y_{[K],j})=U$, the marginal distribution $P_{Q_{i,j}}$ achieves the capacity of $P^{G}_{Y_{i,j}|Q_{i,j}}$.
    By the unique uniform optimum assumption, $P_{Q_{i,j}}$ must be the uniform distribution.
    Since $P_{X_{[K],j}}(\mathcal W)=1$ and $P_{Q_{[K],j}}$ is uniform, the corresponding classical strategy $P_{A_{[K],j}|Q_{[K],j}}$ wins $G$ with probability one.
    This contradicts the fact that $G$ is a pseudo-telepathy game.
    Therefore, we conclude that $I(X_{[K],j} ; Y_{[K],j}) < U$ for all $j$.
    Since the set of product input distributions $\prod_{i\in[K]}P_{X_{i,j}}$ is compact and $I(X_{[K],j} ; Y_{[K],j})$ is continuous in the input distribution, there exists $\delta > 0$ such that 
    \begin{equation}
        I(X_{[K]};Y_{[K]}) \leq U - \delta,
    \end{equation}
    for every product input distribution.
    Together with \eqref{eq:conv_MI_subadd}, we complete the proof.
    }
\end{IEEEproof}

\begin{remark}
    If the inequality in \eqref{eq:conv_gen} is strict when $r=\mathrm{Q}$ and $G$ is a $\mathrm{NS}$-pseudo-telepathy game with $p_w^\mathrm{Q}(G) < 1$, then we can show $\mathcal{C}_s^{\mathrm{Q}} < \mathcal{C}_s^{\mathrm{NS}}$ for the corresponding $G$-game channel.
    However, it is challenging to prove whether the inequality in \eqref{eq:conv_gen} is strict in such a case, based on the similar approach in the last part of the above proof.
    {\rev For $r=C$, Lemma~\ref{lem:C_eq_C_s} reduces the problem to product input distributions, which allows one to derive a contradiction from the absence of a perfect classical winning strategy.
    In contrast, a quantum cooperative encoder may generate correlated question variables, and the conditional distribution $P_{A_{[K]}|Q_{[K]}}$ induced from the code need not be a quantum strategy for the original nonlocal game.}
\end{remark}

\begin{figure*}
        \centering
        \includegraphics[width=0.85\linewidth]{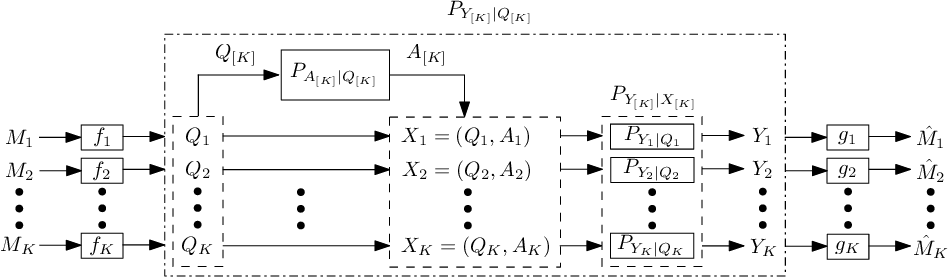}
        \caption{A schematic diagram of the proposed achievability scheme over a single use $(n=1)$ of a game channel $P^G_{Y_{[K]}|X_{[K]}}$, which exploits a winning strategy $P_{A_{[K]}|Q_{[K]}}$ for a pseudo-telepathy game $G$.
        For $n \geq 2$, the part in the dash-dot line is copied $n$ times, i.e., $P_{Y_{[K]}|Q_{[K]}}$ is replaced by $P_{Y_{[K]}^n|Q_{[K]}^n} = P_{Y_{[K]}|Q_{[K]}}^{\otimes n}$.}
        \label{fig:achiev}
\end{figure*}

\subsubsection{Achievability}
We show that the upper bound on the $r$-sum capacity in Proposition~\ref{prop:converse} is achievable if a game $G$ is an $r$-pseudo-telepathy game.
The main idea of the proof is to design an $(n,R_{[K]},r)$-code that incorporates cooperative encoding based on a winning strategy of a game $G$.
Using this code, $P^G_{Y_{[K]}|X_{[K]}}$ splits into parallel channels, i.e., $P^G_{Y_{[K]}|X_{[K]}} = \prod_{i \in [K]} P^G_{Y_i|Q_i}$, and each $i$-th message can be reliably communicated at a rate equal to the capacity of the $i$-th subchannel $P^G_{Y_i|Q_i}$.
\begin{proposition}\label{prop:achiev}
    If $G$ is an $r$-pseudo-telepathy game, then 
    \begin{equation}\label{eq:sum_cap_pseudo_telepathy}
        \mathcal{C}_s^r(P_{Y_{[K]}|X_{[K]}}^{G}) = \sum\limits_{i \in [K]} (\log|\mathcal{Y}_i| - h_i^w).
    \end{equation}
\end{proposition}
\begin{IEEEproof}
    We consider a $(n,R_{[K]},r)$-code depicted in Figure~\ref{fig:achiev}.
    Each of the transmitters first encodes their message with local encoding functions $f_i:\mathcal{M}_i \rightarrow \mathcal{Q}_i^n$.
    Then, for the $j$-th channel use, the transmitters exploit the winning strategy $P_{A_{[K]}|Q_{[K]}} \in \mathcal{E}^r$ of the game $G$ with $Q_{[K],j}$ to generate $A_{[K],j}$, and transmit $X_{[K],j} = (Q_{[K],j} , A_{[K],j})$ to the channel.
    Note that the cooperative encoding $P_{X^n_{[K]}|M_{[K]}}$ constructed in the above is in $\mathcal{E}^r$.
    Since $\mathrm{Pr}(X_{[K],j} \in \mathcal{W})=1$, each of the $j$-th channel use behaves as $K$ parallel channels from $\mathcal{Q}_{[K]}$ to $\mathcal{Y}_{[K]}$, i.e., $P_{Y_{[K],j}|X_{[K],j}} = \prod_{i \in [K]} P_{Y_{i,j}|Q_{i,j}}$.
    Here, $P_{Y_{[K],j}|Q_{[K],j}}$ does not depend on $j \in [n]$.
    Hence, the overall communication scenario is equivalent to the communication over the $n$-times use of parallel channels $\prod_{i \in [K]} P_{Y_i|Q_i}^{\otimes n}$ without cooperation between transmitters.
    Thus, for each $i\in[K]$, the $i$-th transmitter-receiver pair can reliably communicate the message $M_i$ with the rate equal to the point-to-point channel capacity $\mathcal{C}(P_{Y_i|Q_i})$ of the $i$-th subchannel $P_{Y_i|Q_i}$.
    Finally, Lemma~\ref{lem:weak_sym_ch} implies that the sum rate in \eqref{eq:sum_cap_pseudo_telepathy} is achievable.
\end{IEEEproof}

{\rev 
\begin{remark}
    Note that in Proposition~\ref{prop:achiev}, the sum capacity is the same as the mutual information $I(X_{[K]};Y_{[K]})$, where the input distribution $P_{X_{[K]}} (q_{[K]}, a_{[K]}) = P_{Q_{[K]}}(q_{[K]}) P_{A_{[K]}|Q_{[K]}}(a_{[K]}|q_{[K]})$ exploits the winning strategy $P_{A_{[K]}|Q_{[K]}}$ of a pseudo-telepathy game $G$ and $P_{Q_{[K]}}$ is the uniform distribution over $\mathcal{Q}_{[K]}$.
    Since the mutual information is continuous in the input distribution, a quantum or no-signaling advantage in Theorem~\ref{thm:main} persists under sufficiently small deviations from the exact winning strategy $P_{A_{[K]}|Q_{[K]}}$.
\end{remark}
}

\section{Conclusion and Discussion}

{\rev 
We proposed a class of DM-ICs and DM-MACs induced by nonlocal games.
In the winning mode, the channel inputs satisfy an alignment condition: residual cross-link interference is removed, and the channel decomposes into parallel weakly symmetric subchannels whose capacity-achieving input distributions are unique.
Outside the winning mode, the channel has strictly larger conditional output uncertainty.
We showed that, when the underlying game is a quantum or no-signaling pseudo-telepathy game, quantum or no-signaling cooperation between transmitters strictly increases the sum capacity compared with classical cooperation.
The examples based on the uniform random noise and the cyclic nearest-neighbor confusion channels illustrate that the proposed class of channels covers models beyond the constructions used in earlier works.

Our theorem establishes a strict separation for the proposed finite-alphabet $G$-game channels.
However, two questions remain open: whether the exact magnitude of the classical-to-quantum gap can be efficiently computed, and whether analogous separations can be proved for more standard physical channel models.

For the first point, it is necessary to quantify the sum capacity without cooperation.
As one approach, the generalized Blahut-Arimoto algorithm \cite{rezaeianComputationTotalCapacity2004}, proposed for computing the classical sum capacity of DM-MACs (which serves as an upper bound on the sum capacities of DM-ICs), could be employed.
However, although this algorithm can evaluate the sum capacities of certain DM-MACs, it generally provides only lower bounds on the sum capacities \cite{buhlerNoteCapacityComputation2011}.
Subsequently, \cite{seshadriSeparationCorrelationAssistedSum2023} proposed an algorithm that computes the sum capacities of two-transmitter DM-MACs within a given additive precision in quasi-polynomial time.
As a future direction, it would also be interesting to investigate efficient algorithms for computing the sum capacity in scenarios with more than two transmitters.

For the second point, it would be desirable to determine whether quantum cooperation can enhance the sum capacity of practical channels, such as Gaussian fading ICs and MACs.
Theorem~\ref{thm:main} does not directly apply to Gaussian fading ICs and MACs, since the $G$-game channels considered in this work are formulated for discrete channels.
However, the idea of weaker interference in the winning mode would be generalized to Gaussian channels with fading.
Based on this idea, we conjecture that there is a quantum advantage in those channels when the channel gains resemble the structure of a $G$-game channel.
}

Another research direction would be to derive a single-letter characterization of the capacity region of general DM-ICs or DM-MACs with cooperation between transmitters.
In this context, \cite{peregMultipleAccessChannelEntangled2025} derived a multi-letter characterization of the capacity region of a DM-MAC, and \cite{hawellekInterferenceChannelEntangled2025} proposed multi-letter inner and outer bounds for the capacity region of a DM-IC.
Compared to single-letter characterizations, multi-letter characterizations have the limitation that it is not clear how to compute them \cite{elgamalNetworkInformationTheory2011}.
A single-letter characterization would help characterize the set of all channels in which cooperation enlarges the capacity region.

\section*{Acknowledgments}
This work was supported in part by Basic Science Research Program through the National Research Foundation of Korea (NRF) funded by the Ministry of Education (No. RS-2024-00452156), in part by the Institute of Information \& Communications Technology Planning \& Evaluation (IITP) grant funded by the Korea government (MSIT) (RS-2023-00229524, RS-2025-02304540, RS-2025-25464876, RS-2025-25464616), in part by the NRF grant funded by the Korea government (MSIT) (No. RS-2025-00561467), and in part by the KASA (Korea AeroSpace Administration) under Grant 2022M1A3C2069728, Future Space Education Center.

\bibliographystyle{IEEEtran}
\bibliography{ref}

\begin{IEEEbiography}[{\includegraphics[width=1in,height=1.25in,clip,keepaspectratio]{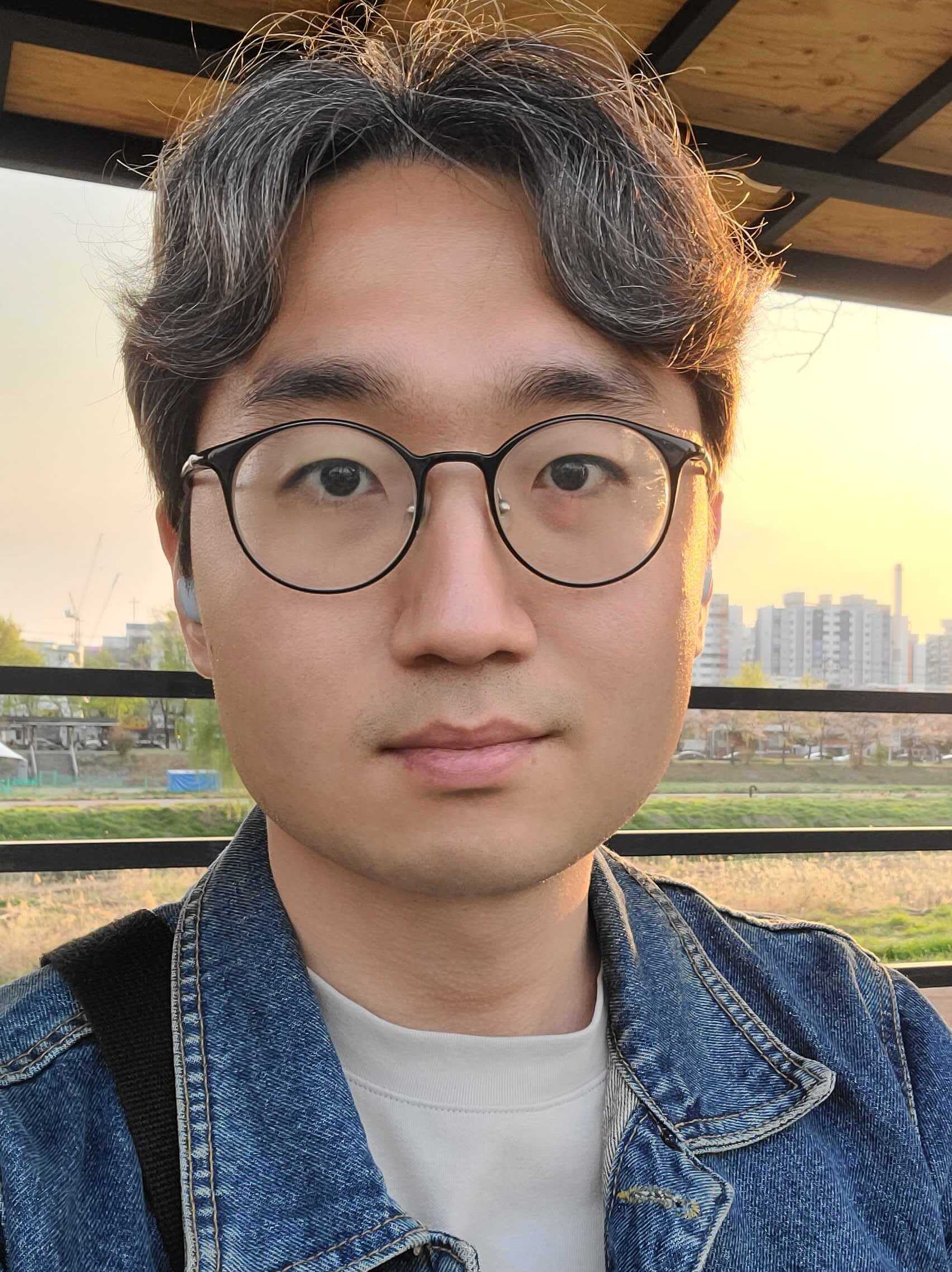}}]{Seung-Hyun Nam} (Member, IEEE) received the B.S. and M.S. degrees in Electrical Engineering from Pohang University of Science and Technology (POSTECH), Pohang, South Korea, in 2018 and 2020, respectively, and the Ph.D. degree in Electrical Engineering from Korea Advanced Institute of Science and Technology (KAIST), Daejeon, South Korea, in 2024.
He is currently a post-doctoral researcher with the Information \& Electronics Research Institute, KAIST.
His research interests include (both classical and quantum) information theory, statistical inference, differential privacy, and information-theoretic security.
\end{IEEEbiography}

\begin{IEEEbiography}[{\includegraphics[width=1in,height=1.25in,clip,keepaspectratio]{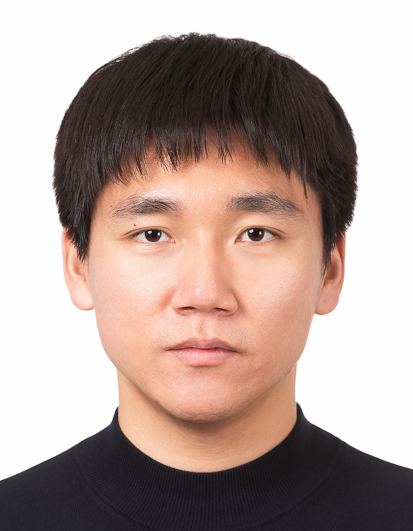}}]{Hyun-Young Park} (Member, IEEE) received the B.S. (valedictorian) and Ph.D. degrees in Electrical Engineering from Korea Advanced Institute of Science and Technology (KAIST), Daejeon, South Korea, in 2021 and 2026, respectively.
He is currently a post-doctoral researcher with the School of Electrical Engineering, KAIST. His research interests include information theory, differential privacy, and quantum information theory.
\end{IEEEbiography}

\begin{IEEEbiography}[{\includegraphics[width=1in,height=1.25in,clip,keepaspectratio]{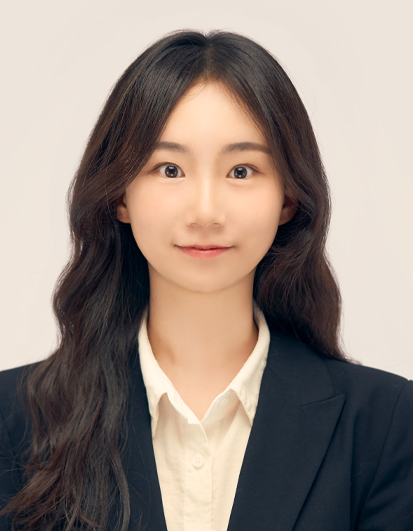}}]{Jiyoung Yun} received the B.S. and M.S. degrees in Applied Mathematics from Hanyang University, and the Ph.D. degree in Electrical Engineering from the Korea Advanced Institute of Science and Technology (KAIST). She is currently a postdoctoral researcher at KAIST. Her research interests include quantum information theory, network quantum information theory, classical and quantum channel capacities, and entanglement-assisted communication.
\end{IEEEbiography}

\begin{IEEEbiography}
[{\includegraphics[width=1in,height=1.25in,clip,keepaspectratio]{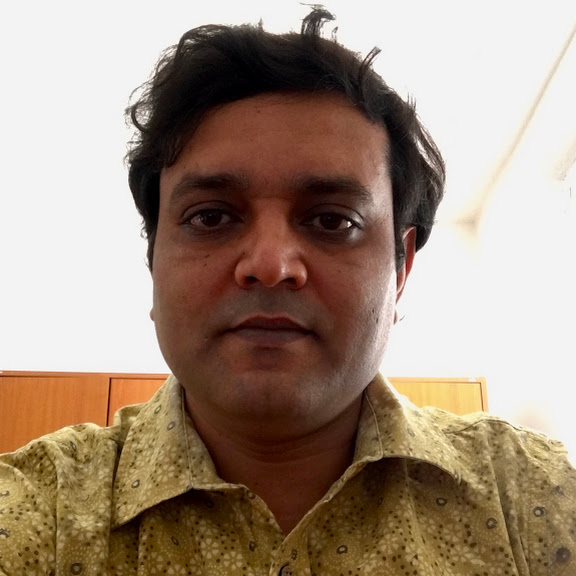}}]{Ashutosh Rai} did his Ph.D. work [2009-2013] in theoretical physics from the S. N. Bose National Center for Basic Sciences (University of Calcutta, Kolkata, India). He has worked as a researcher at the University of Latvia (Latvia), International Institute of Physics (Brazil), and Slovak Academy of Sciences (Slovakia). He is currently working at the School of Electrical Engineering, Korea Advanced Institute of Science and Technology (KAIST). His research interest cover topics in foundation of quantum mechanics and quantum information theory. 
\end{IEEEbiography}

\begin{IEEEbiography}[{\includegraphics[width=1in,height=1.25in,clip,keepaspectratio]{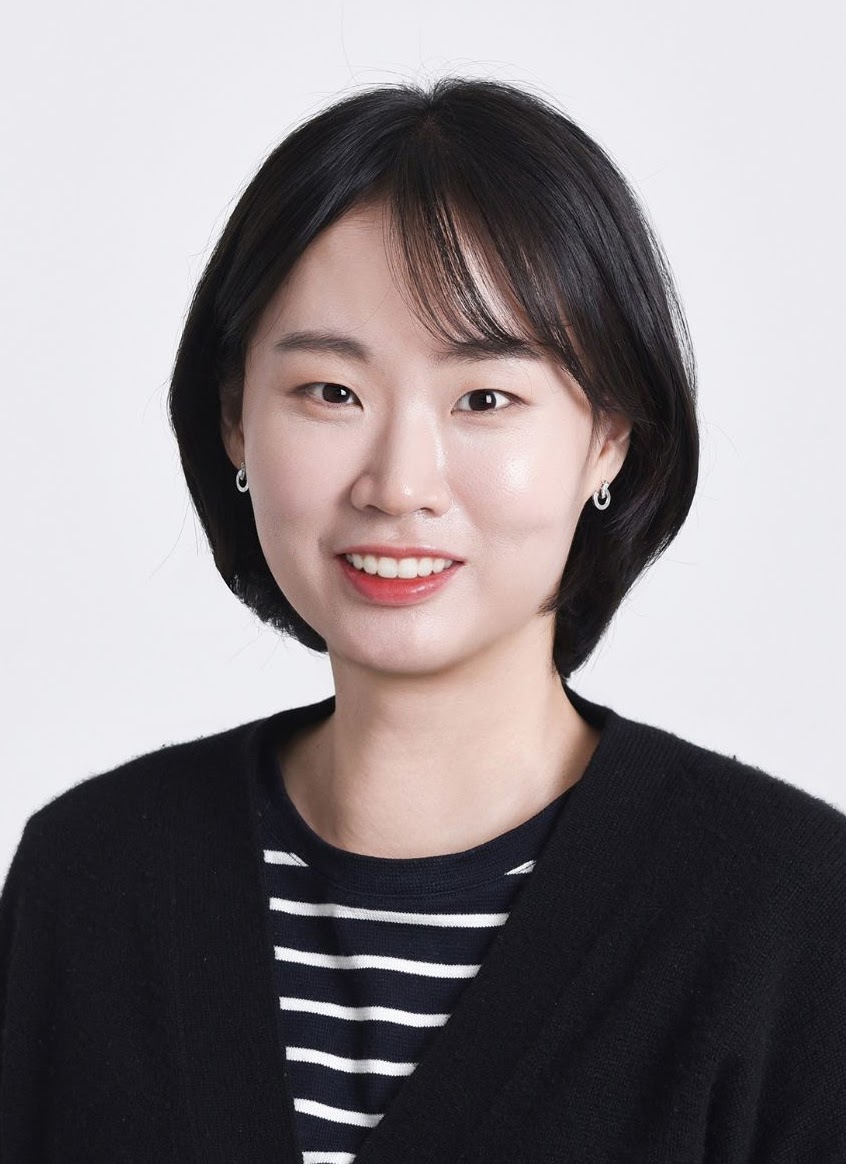}}]
{Si-Hyeon Lee} (Senior Member, IEEE) received the B.S. (summa cum laude) and Ph.D. degrees in electrical engineering from the Korea Advanced Institute of Science and Technology (KAIST), Daejeon, South Korea, in 2007 and 2013, respectively. She is currently an Associate Professor with the School of Electrical Engineering, KAIST. She was a Postdoctoral Fellow with the Department of Electrical and Computer Engineering, University of Toronto, Toronto, Canada, from 2014 to 2016, and an Assistant Professor with the Department of Electrical Engineering, Pohang University of Science and Technology (POSTECH), Pohang, South Korea, from 2017 to 2020. Her research interests include information theory, wireless communications, statistical inference, and machine learning. 
She was an IEEE Information Theory Society Distinguished Lecturer (2024-2025) and a TPC Co-Chair of the IEEE Information Theory Workshop 2024. She served as  a Guest Editor for the {\scshape IEEE Journal on Selected Areas in Communications} (Special Issue on Secure Communication, Sensing, and Computation in Future Intelligent Wireless Networks). She is currently an Associate Editor for the {\scshape IEEE Transactions on Information Theory} and the {\scshape IEEE Transactions on Communications}.
\end{IEEEbiography}

\begin{IEEEbiography}[{\includegraphics[width=1in,height=1.25in,clip,keepaspectratio]{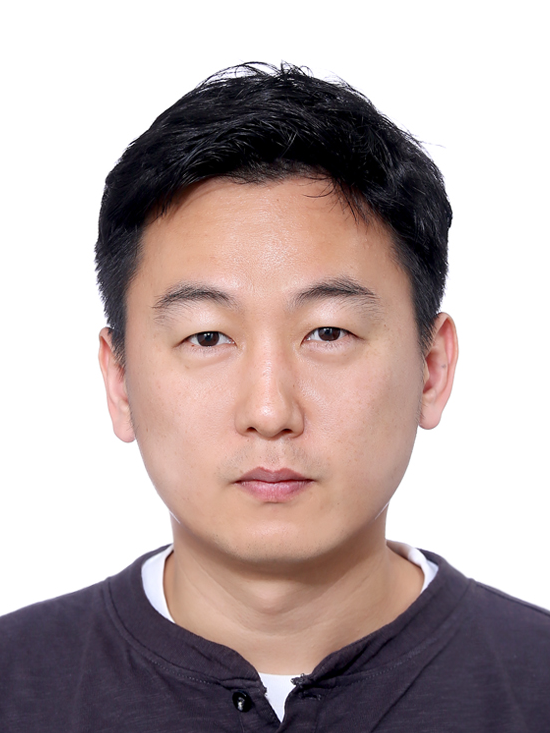}}]{Joonwoo Bae} (Member, IEEE) obtained a Ph.D. in Theoretical Physics from Universitat de Barcelona \& ICFO-Institute of Photonic Sciences, Barcelona in 2007. He has worked at the Korea Institute for Advanced Study (KIAS), Centre for
Quantum Technologies (CQT) in Singapore, the ICFO, Freiburg Institute for Advanced Studies (FRIAS) as a Junior Fellow, and Hanyang University. He is currently with the School of Electrical Engineering, Korea Advanced Institute of Science and Technology (KAIST). His research interests include secure quantum communication, entanglement applications, open quantum systems, quantum foundations, and their practical applications.
\end{IEEEbiography}

\vfill

\end{document}